\documentclass[epj]{svjour}
\usepackage{latexsym}
\usepackage{graphics}
\usepackage[latin1]{inputenc}
\usepackage{rotating}
\usepackage{amsmath}
\begin{document}
\renewcommand{\textfraction}{0.00000000001}
\renewcommand{\floatpagefraction}{1.0}
\title{Measurement of the beam-helicity asymmetry $I^{\odot}$ in the  
photoproduction of $\pi^0\pi^{\pm}$-pairs off protons and off neutrons}
\author{
M.~Oberle\inst{1},
J.~Ahrens\inst{2},		
J.R.M.~Annand\inst{3},	   
H.J.~Arends\inst{2},
K.~Bantawa\inst{4},
P.A.~Bartolome\inst{2},	    
R.~Beck\inst{5},	 
V.~Bekrenev\inst{6},
H.~Bergh\"auser\inst{7},	       
A.~Braghieri\inst{8},  	 
D.~Branford\inst{9},	     
W.J.~Briscoe\inst{10},	      
J.~Brudvik\inst{11},		
S.~Cherepnya\inst{12},	   
B.~Demissie\inst{10},
M.~Dieterle\inst{1},		  
E.J.~Downie\inst{2,3,10},	    
P.~Drexler\inst{7}, 
L.V. Fil'kov\inst{12}, 
A.~Fix\inst{13},
D.I.~Glazier\inst{9},	     
E.~Heid\inst{2},
D.~Hornidge\inst{14}, 
D.~Howdle\inst{3}, 
G.M.~Huber\inst{15},		
O.~Jahn\inst{2},
I.~Jaegle\inst{1},
T.C.~Jude\inst{9},
A. K{\"a}ser\inst{1},  	      
V.L.~Kashevarov\inst{12,2},
I. Keshelashvili\inst{1},	  
R.~Kondratiev\inst{16},	  
M.~Korolija\inst{17},  
S.P.~Kruglov\inst{6}, 
B.~Krusche\inst{1},
A.~Kulbardis\inst{6},
V.~Lisin\inst{16},		  
K.~Livingston\inst{3},	   
I.J.D.~MacGregor\inst{3},	   
Y.~Maghrbi\inst{1},
J.~Mancell\inst{3},  
D.M.~Manley\inst{4}, 
Z.~Marinides\inst{10},	      
M.~Martinez\inst{2},
J.C.~McGeorge\inst{3}, 
E.~McNicoll\inst{3}, 	 
D.~Mekterovic\inst{17},	  
V.~Metag\inst{7},
S.~Micanovic\inst{17},
D.G.~Middleton\inst{14},
A.~Mushkarenkov\inst{8},		
B.M.K.~Nefkens\inst{11}, 	
A.~Nikolaev\inst{5},   
R.~Novotny\inst{7},
M.~Ostrick\inst{2},
B.~Oussena\inst{2,10},
P.~Pedroni\inst{8}, 
F.~Pheron\inst{1},	       
A.~Polonski\inst{16},  	  
S.N.~Prakhov\inst{11}, 
J.~Robinson\inst{3}, 		   
G.~Rosner\inst{3},		   
T.~Rostomyan\inst{1,8},	       
S.~Schumann\inst{2},
M.H.~Sikora\inst{9},	     
D.I.~Sober\inst{18}, 	      
A.~Starostin\inst{11},		
I.~Supek\inst{17},		  
M.~Thiel\inst{7},	    
A.~Thomas\inst{2},		 
M.~Unverzagt\inst{2,5},			  
D.P.~Watts\inst{9},
D.~Werthm\"uller\inst{1},
L.~Witthauer\inst{1}, 
F.~Zehr\inst{1}
\newline(The Crystal Ball at MAMI, TAPS, and A2 Collaborations)
\mail{B. Krusche, Klingelbergstrasse 82, CH-4056 Basel, Switzerland,
\email{Bernd.Krusche@unibas.ch}}
}

\institute{Department of Physics, University of Basel, CH-4056 Basel, Switzerland
  \and Institut f\"ur Kernphysik, University of Mainz, D-55099 Mainz, Germany
  \and School of Physics and Astronomy, University of Glasgow, G12 8QQ, United Kingdom
  \and Kent State University, Kent, Ohio 44242, USA
  \and Helmholtz-Institut f\"ur Strahlen- und Kernphysik, University of Bonn, D-53115 Bonn, Germany
  \and Petersburg Nuclear Physics Institute, RU-188300 Gatchina, Russia
  \and II. Physikalisches Institut, University of Giessen, D-35392 Giessen, Germany
  \and INFN Sezione di Pavia, I-27100 Pavia, Pavia, Italy
  \and School of Physics, University of Edinburgh, Edinburgh EH9 3JZ, United Kingdom
  \and Center for Nuclear Studies, The George Washington University, Washington, DC 20052, USA
  \and University of California Los Angeles, Los Angeles, California 90095-1547, USA
  \and Lebedev Physical Institute, RU-119991 Moscow, Russia
  \and Laboratory of Mathematical Physics, Tomsk Polytechnic University, Tomsk, Russia
  \and Mount Allison University, Sackville, New Brunswick E4L1E6, Canada
  \and University of Regina, Regina, SK S4S-0A2 Canada
  \and Institute for Nuclear Research, RU-125047 Moscow, Russia
  \and Rudjer Boskovic Institute, HR-10000 Zagreb, Croatia
  \and The Catholic University of America, Washington, DC 20064, USA
}
\authorrunning{M. Oberle et al.}
\titlerunning{Beam-helicity asymmetries for $\pi^0\pi^{\pm}$-pairs}
	
\abstract
{Beam-helicity asymmetries have been measured at the MAMI accelerator in Mainz 
for the photoproduction of mixed-charge pion pairs in the reactions 
$\boldsymbol{\gamma}p\rightarrow n\pi^0\pi^+$ off free protons and 
$\boldsymbol{\gamma}d\rightarrow (p)p\pi^0\pi^-$ 
and $\boldsymbol{\gamma}d\rightarrow (n)n\pi^0\pi^+$ off quasi-free nucleons 
bound in the deuteron for incident photon energies up to 1.4 GeV. 
Circularly polarized photons were produced from bremsstrahlung of longitudinally 
polarized electrons and tagged with the Glasgow-Mainz magnetic spectrometer. The charged 
pions, recoil protons, recoil neutrons, and decay photons from $\pi^0$ mesons
were detected in the 4$\pi$ electromagnetic calorimeter composed of the Crystal 
Ball and TAPS detectors. Using a complete kinematic reconstruction of the final 
state, excellent agreement was found between the results for free and quasi-free 
protons, suggesting that the quasi-free neutron results are also a close approximation
of the free-neutron asymmetries. A comparison of the results to the predictions of
the Two-Pion-MAID reaction model shows that the reaction mechanisms are still not
well understood, in particular at low incident photon energies in the second 
nucleon-resonance region.
} 
\maketitle

\section{Introduction}
The excitation spectrum of nucleons is a much discussed topic because it is
closely related to the fundamental properties of the strong interaction
in the non-perturbative range. The apparently unsatisfactory match \cite{Krusche_03}
between model predictions based on Quantum Chromodynamics `inspired' quark models 
and the experimental database for excited nucleon states has motivated many recent 
efforts in experiment and also in theory development. Recent progress for 
the latter came mostly from the application of the Dyson-Schwinger equation to the QCD
Lagrangian (see e.g. \cite{Chen_12a,Eichmann_12,Aznauryan_13}) and  
from the advances in lattice gauge calculations and their combination 
with the methods of chiral perturbation theory for the extrapolation to physical 
quark masses. First unquenched lattice results, which recently became available  
\cite{Edwards_11}, basically `re-discovered' the SU(6)$\otimes$O(3) 
excitation structure of the nucleon with a level counting consistent with the standard 
non-relativistic quark model. However, one should keep in mind that these calculations 
are still at a very early stage. On the experimental side, over the last decade,
much effort has been made to overcome the limitations in the available database,
which was dominated by the results from pion scattering on nucleons
and thus biased against nucleon resonances with small couplings to $N\pi$. 
Due to the advances in accelerator and detector technology, photoproduction of mesons 
has become a prime tool in this research.

It was soon realized that sequential decays involving intermediate excited states 
play an essential role, especially for higher lying resonances.
This is in analogy to nuclear physics, where a restriction to the
ground-state decays of excited states would have resulted in a very limited
picture of nuclear structure, missing fundamentally important features like 
collective rotational or vibrational bands. Since excited nucleon states
decay almost exclusively via meson emission, the only possibility
is to study reactions with meson multiplicity larger than one in the final 
state. Therefore, such reactions have attracted much interest in recent years. 
In particular, the production of pseudoscalar meson pairs,
mostly pion pairs but also pion eta pairs, has been experimentally studied.
Special attention was paid to neutral mesons ($\pi^0\pi^0$ and $\pi^0\eta$,
see 
\cite{Haerter_97,Zabrodin_97,Zabrodin_99,Wolf_00,Kleber_00,Kotulla_04,Sarantsev_08,Thoma_08,Zehr_12,Oberle_13,Kashevarov_12,Ajaka_08,Horn_08a,Horn_08b,Kashevarov_09,Kashevarov_10} 
for recent results). The reason is that non-resonant background terms are more important 
for charged pions since the incident photon can couple directly to them. 

However, the measurement of different charge combinations of the pion pairs and also 
measurements of their production off both protons and neutrons is mandatory for an isospin
decomposition of the reaction, helping to identify contributions from $N^{\star}$ and 
$\Delta^{\star}$ resonances. Furthermore, reactions with at least one charged pion in the final 
state should allow the investigation of contributions from resonance decays by emission
of the $\rho$ meson. The $\rho^{\pm}$ meson decays to $\pi^0\pi^{\pm}$ pairs while the 
$\rho^0$ meson cannot decay to $\pi^0\pi^0$ pairs but only to $\pi^+\pi^-$ pairs.

\begin{figure*}[htb]
\resizebox{1.\textwidth}{!}{%
  \includegraphics{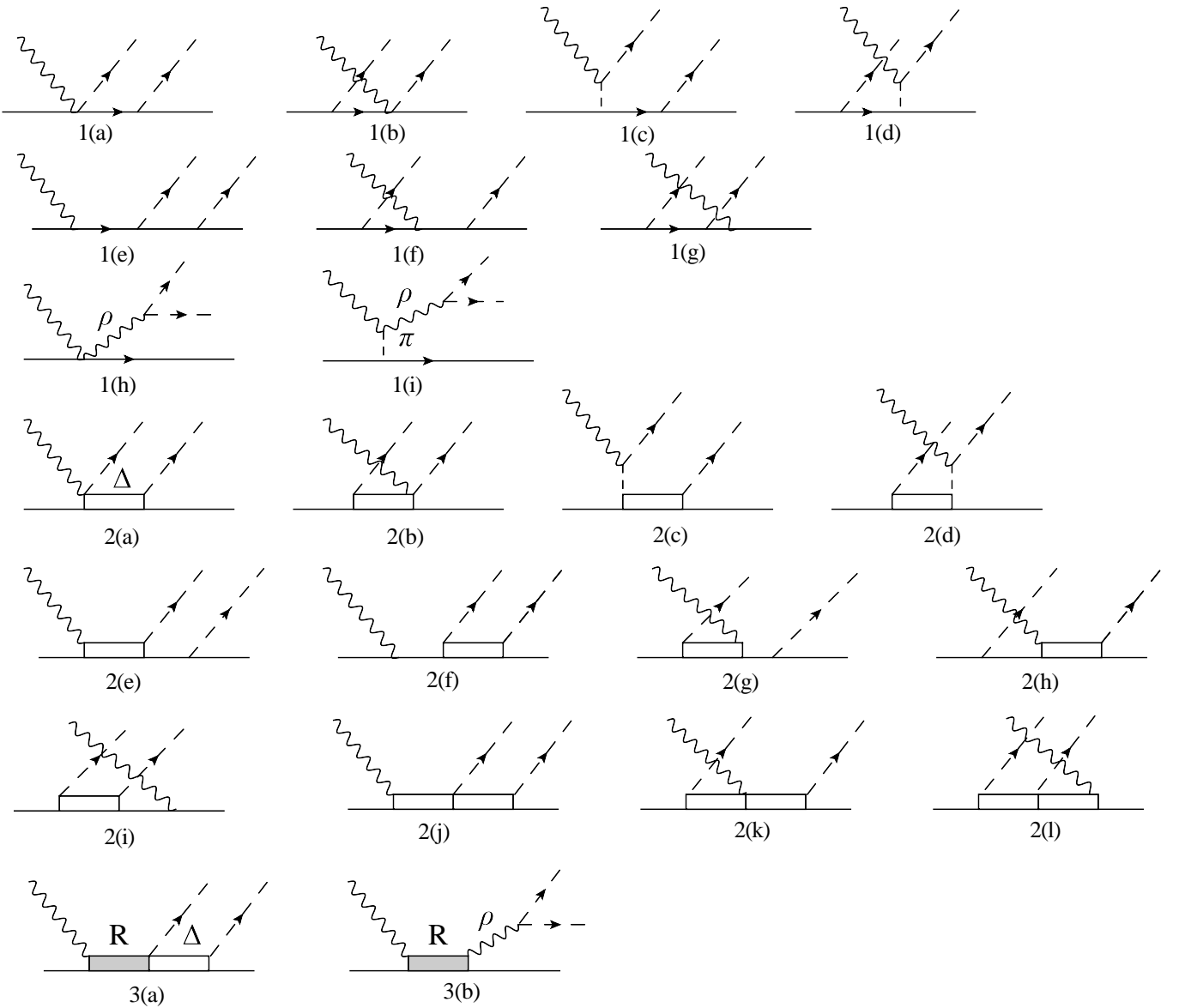}
}
\caption{Contributions to photoproduction of meson pairs at low incident 
photon energy. Shown are the diagrams that are considered in the Two-Pion-MAID
model \cite{Fix_05}. Diagrams 1(a) - 1(i) correspond to nucleon Born, 
pion-pole (c-d,i), and $\rho$-Kroll Ruderman (h) background terms. 
Diagrams 2(a) - 2(l) represent similar background terms involving the $\Delta(1232)$ 
state. The s-channel diagram 3(a) represents sequential decays of higher lying 
resonances via the $\Delta$(1232) intermediate state and 3(b) the direct decay
of resonances to the nucleon ground state via emission of $\rho$-mesons.  
}
\label{fig:diag}       
\end{figure*}

Predictions for many different observables for all possible isospin channels are 
available from the Two-Pion-MAID reaction model by Fix and Arenh\"ovel \cite{Fix_05}.
The model is based on an effective Lagrangian approach with Born and resonance diagrams
at the tree level. They are summarized in Fig.~\ref{fig:diag}.
The most interesting diagrams for the investigation of nucleon resonances are 3(a) and
3(b) which include all $s$-channel Breit-Wigner resonances $R$ with $J\leq 5/2$ for
which all parameters like $N\gamma$ coupling and partial $R\rightarrow\pi\Delta$ and
$R\rightarrow N\rho$ decay widths were taken from the Particle Data Group. 
Diagram 3(a) corresponds to sequential resonance decays where a higher lying excited 
state ($N^{\star}$ or $\Delta^{\star}$) decays to the $\Delta$(1232) resonance.
Diagram 3(b) corresponds to the direct groundstate $N\rho$-decays of nucleon resonances.
Non-resonant background contributions arise from the nucleon-nucleon and nucleon-$\Delta$ 
Born-terms and from pion-pole terms.  Non-re\-so\-nant background contributes also for 
the charged $\rho$-meson for example from the $\rho$-Kroll-Ruderman term (diagram 1(h)) 
and the pion-pole term (1(i)). The contributions of the non-resonant backgrounds for 
charged mesons may be substantial. This is already reflected \cite{Krusche_03} in the 
absolute magnitude 
($\sigma\approx$~10~$\mu$b for $\pi^0\pi^0$,  $\sigma\approx$~50~$\mu$b for $\pi^0\pi^+$, 
$\sigma\approx$~75~$\mu$b for $\pi^+\pi^-$) of the total cross section for the different 
charge states at the second nucleon resonance region composed of the $P_{11}$(1440), 
$D_{13}$(1520), and $S_{11}$(1535) state.  

Production of $\pi^+\pi^-$ pairs has been studied by electron scattering at the CLAS 
facility at JLab \cite{Ripani_03,Fedotov_09}. The results
have been interpreted in the framework of a phenomenological meson-baryon reaction model 
and used to extract electrocouplings for the $P_{11}$(1440) and $D_{13}$(1525)
resonances \cite{Mokeev_09,Mokeev_12}. 

The first precise measurements of the total cross section and the invariant-mass 
distributions of pion-pion and pion-nucleon pairs for the $\pi^0\pi^+$ final state for 
photoproduction off the proton in the second resonance region was done with the
DAPHNE \cite{Braghieri_95} and TAPS \cite{Langgaertner_01} experiments at the MAMI accelerator 
in Mainz. They revealed strong discrepancies with all available model predictions 
\cite{Krusche_03,Langgaertner_01}. It was then suggested that the main effect was 
caused by the neglect of $\rho$ contributions in the models (due to the relatively large 
mass of this meson, $\rho$ contributions had been neglected in the second resonance 
region.) Including such terms significantly improved the agreement 
between reaction models and experimental results \cite{Ochi_97,Nacher_01}. Particularly, the 
invariant-mass distributions measured by DAPHNE for ${\gamma} n\rightarrow p\pi^-\pi^0$
\cite{Zabrodin_99} in quasi-free kinematics off neutrons bound in the deuteron and by 
TAPS for $\gamma p\rightarrow n\pi^0\pi^+$ \cite{Langgaertner_01} supported this 
interpretation. The latter results were taken as evidence for a significant
contribution from the $D_{13}(1520)\rightarrow N\rho$ decay to the photoproduction of
mixed charge pairs \cite{Langgaertner_01}, which was subsequently also confirmed
for the $\pi^+\pi^-$ final state in electroproduction \cite{Mokeev_12}.

The discussions of the $\rho$-meson contribution were based on the shape difference of 
the pion-pion invariant mass distributions for the $\pi^0\pi^0$ and $\pi^0\pi^{\pm}$ final 
states and the comparison to the results from reaction models, in particular to the Valencia 
model \cite{Nacher_01}. 
However, the problem is much more complicated. Photoproduction of pseudo-scalar meson pairs
off nucleons involves eight complex amplitudes \cite{Roberts_05} as function of five kinematic
variables (for example two Lorentz invariants and three angles). The measurement of eight
independent observables would be needed just to extract the magnitude of all amplitudes in a 
unique way (not even considering ambiguities arising from finite statistical precision of the 
data). Fixing in addition the phases would require the measurement of 15 observables. It is
thus evident that the analysis of differential cross section data alone cannot solve the
problem. Analyses based on such a limited data set will always strongly depend on the
model assumptions. A more profound analysis requires the measurement of further observables, 
exploring polarization degrees of freedom. A fully complete measurement appears unrealistic 
due to the huge effort needed, but already the measurement of at least some polarization 
observables can provide valuable constraints for the reaction models.

The reaction $\boldsymbol{\gamma} \vec{p}\rightarrow n\pi^0\pi^+$ has been measured with the 
DAPHNE detector at MAMI with a circularly polarized photon beam and a longitudinally polarized 
target for incident photon energies up to 800 MeV \cite{Ahrens_03}. The results have been used 
to split the cross section into the $\sigma_{3/2}$ (photon and proton spins parallel) and 
$\sigma_{1/2}$ (spins antiparallel) components. The result shows a dominance of the 
$\sigma_{3/2}$ part in the second resonance region. This would be in line with contributions 
from the $D_{13}$(1520) resonance, either via the sequential 
$D_{13}\rightarrow\Delta(1232)\pi\rightarrow N\pi\pi$ decay chain and/or the direct 
$D_{13}\rightarrow N\rho$ decay. The Valencia model with the $\rho$-terms and an additional 
contribution from the $D_{33}$(1700) resonance \cite{Nacher_01,Nacher_02} agreed well with the 
$\sigma_{3/2}$ component, but somewhat underestimated $\sigma_{1/2}$.

At this point another polarization observable moved into the focus, namely the beam
helicity asymmetry $I^{\odot}$, measured with circularly polarized photon beams
and unpolarized targets. Reaction models \cite{Roca_05} had predicted a large sensitivity
to small contributions via interference terms. The first measurement of this observable 
for the $p\pi^+\pi^-$ final state at JLab \cite{Strauch_05} revealed severe deficiencies in all 
reaction models for this observable. A subsequent measurement \cite{Krambrich_09} of it for 
all possible final states off the proton target ($p\pi^+\pi^-$, $n\pi^+\pi^0$, $p\pi^0\pi^0$) 
in the second resonance region confirmed the results for the doubly charged pion pairs, and
showed similar or even worse problems for the mixed-charge final state, while only the
measured asymmetries for the  $2\pi^0$ final state were reasonably well reproduced by
some reaction models. In the meantime, asymmetries for double $\pi^0$ production have
been measured off free protons and quasi-free protons and neutrons bound in the deuteron
up to incident photon energies of 1.4 GeV \cite{Oberle_13}. Reaction-model results
are in reasonable agreement with the measured asymmetries for the reaction off the proton,
but less so for the neutron target. A surprising result was that the Valencia model
\cite{Nacher_01,Nacher_02,Roca_05} failed for $I^{\odot}$ in all isospin channels, although 
it had reasonably well reproduced all other observables measured so far for the different 
final states (total cross sections, invariant-mass distributions, $\sigma_{3/2}$ - $\sigma_{1/2}$ 
decomposition of the cross sections).   

The present work reports the results from the measurement of beam-helicity asymmetries
in photoproduction of $\pi^0\pi^{\pm}$ pairs off free protons, quasi-free protons, and 
neutrons for incident photon energies up to 1.4 GeV. Measurements off the neutron can only be
done in quasi-free kinematics off neutrons bound in light nuclei, specifically the deuteron.
This is complicated by the nuclear Fermi motion and possible final-state interaction (FSI)
effects, but much progress has recently been made in the analysis and interpretation of 
such reactions \cite{Krusche_11}.  

\section{Beam-helicity asymmetries}
\label{sec:I_intro}

\begin{figure}[thb]
\resizebox{0.50\textwidth}{!}{%
  \includegraphics{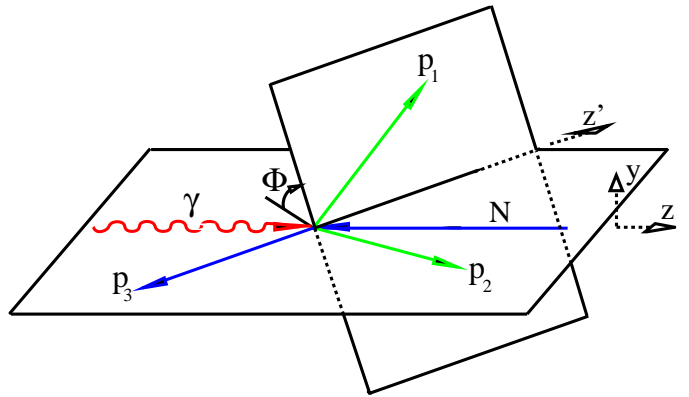}
}
\caption{Vector and angle definitions in the cm system of incident photon ($\gamma$) and 
initial-state participant nucleon $N$. Particles $p_1$, $p_2$, and $p_3$ are some
permutation of the final-state participant nucleon $N'$ and the two pions ($\pi^0$, $\pi^{\pm}$),
depending on the type of the asymmetry (see text). One plane is defined by the momentum
of the incident photon $\vec{k}$ and the momentum of particle $p_3$, the other by the
momenta of particles $p_1$ and $p_2$ (all momenta in the photon -nucleon cm system).
$\Phi$ is the angle between the planes. For the choice $p_3 = N'$, $p_1 = \pi^{\pm}$,
$p_2 = \pi^0$ the planes are the usual reaction and production planes as defined in 
Refs.~\cite{Krambrich_09,Roca_05}. 
}
\label{fig:def}       
\end{figure}

Beam-helicity asymmetries $I^{\odot}$ can be measured for three-body final states 
like $N\pi\pi$ with circularly polarized photons and unpolarized targets. This observable is 
defined by:
\begin{equation}
I^{\odot}(\Phi)=
	        \frac{d\sigma^{+}-d\sigma^{-}}{d\sigma^{+}+d\sigma^{-}}
	       =\frac{1}{P_{\gamma}}
                \frac{N^{+}-N^{-}}{N^{+}+N^{-}}\;,
\label{eq:circ}
\end{equation}
where $d\sigma^{\pm}$ are the differential cross sections for each of the two photon helicity 
states, and $P_{\gamma}$ is the degree of circular polarization of the photons. The angle $\Phi$
can be defined in different ways in the cm system of the incident photon and the initial state
nucleon. This is illustrated in Fig. \ref{fig:def}. Two planes are spanned by the incident photon,
the recoil nucleon, and the two pions and $\Phi$ is the angle between them. 

Beam helicity asymmetries are particularly robust with respect to false asymmetries introduced
by the experimental setup. First of all such effects cancel in the ratio Eq.~\ref{eq:circ}. 
Furthermore, the angle $\Phi$ does not correspond  to a specific azimuthal direction in the 
laboratory system. In the laboratory, the whole system shown in Fig.\ref{fig:def}
can be arbitrarily rotated around the beam axis, so that for each value of $\Phi$ the experiment
averages automatically over all azimuthal orientations in the detector system.  
Any effects from the dependence of the experimental detection efficiency on the azimuthal angle 
in the detector frame are thus eliminated. 

For the most basic version, defined as in \cite{Krambrich_09,Roca_05}, we choose the outgoing 
recoil nucleon as particle $p_3$, spanning together with the photon the reaction plane, while the 
two pions are chosen as particles $p_1$, $p_2$ and span the production plane. The definition of the 
angle $\Phi$ depends then still on the ordering of the pions, for which we can use different prescriptions. 
For the non-identical pions in the $N\pi^0\pi^{\pm}$ final state the most natural 
ordering is by their charge. For this we use the same convention as in \cite{Krambrich_09}, 
i.e., $p_1 = \pi^{\pm}$ is the charged pion and $p_2 = \pi^0$ the neutral one (this analysis is 
called `charge ordered'), the corresponding asymmetry is denoted as $I^{\odot}_{1c}(\Phi_{1c})$. 
We can also order them by the reaction kinematics, which is the only possibility for identical pions. 
For this we use the same condition as for the doubly neutral pairs in \cite{Oberle_13}, where the 
pion with the larger pion-nucleon invariant mass is chosen as $p_1$ 
\begin{equation}
m(\pi_1,N')\ge m(\pi_2,N') 
\label{eq:mmcon}
\end{equation}
and the results, $I^{\odot}_{1m}(\Phi_{1m})$, are labeled `mass ordered'.

There are actually further asymmetries, which have not been considered in previous analyses. 
They arise when we choose one of the pions as $p_3$. We use the following definitions.
The asymmetry $I^{\odot}_{2c}(\Phi_{2c})$ corresponds to the choice  
$(p_1,p_2,p_3) = (\pi^0,N',\pi^{\pm})$ and $I^{\odot}_{3c}(\Phi_{3c})$ to
$(p_1,p_2,p_3) = (\pi^{\pm},N',\pi^{0})$.

Due to parity conservation all asymmetries must obey the condition: 
\begin{equation}
I^{\odot}(\Phi)=-I^{\odot}(2\pi-\Phi) \nonumber .
\label{eq:sym1}
\end{equation}

For the extraction of $I^{\odot}(\Phi,\Theta_{\pi_1},\Theta_{\pi_2},...)$ in a limited region 
of kinematics, the differential cross sections $d\sigma^{\pm}$ can be  
replaced by the respective count rates $N^{\pm}$ (right hand side of Eq.~\ref{eq:circ}), since 
all normalization factors cancel in the ratio. For angle-integrated asymmetries, efficiency-weighted 
count rates $N/\epsilon$ should be used in the integration.   

Due to their symmetry properties, the $I^{\odot}$ can be expanded in sine series
\begin{equation}
I^{\odot}(\Phi)=\sum_{n=1}^{\infty}A_{n}\mbox{sin}(n\Phi)
\label{eq:coeff}
\end{equation}
which can be fitted to the data. The coefficients with even numbers must be identical 
(within uncertainties) for the asymmetries  $I^{\odot}_{1c}$ and  $I^{\odot}_{1m}$
(`charge' or `mass' ordering of the pions in one plane), while the odd coefficients 
depend on the ordering (and have to vanish for `random' ordering).
 
\section{Experimental setup}
The experiments were performed at the tagged photon facility of the Mainz Microtron accelerator MAMI
\cite{Herminghaus_83,Kaiser_08}. Longitudinally polarized electron beams with energies of 
$\approx$1.5 GeV (see Table \ref{tab:beam} for details) were used to produce bremsstrahlung
photons in a copper radiator of 10 $\mu$m thickness, which were tagged with the upgraded 
Glasgow magnetic spectrometer \cite{Anthony_91,Hall_96,McGeorge_08}. The typical bin width 
for the photon beam energy (4 MeV) was defined by the geometrical size of the plastic 
scintillators in the focal plane detector of the tagger. 
The polarization degree of the electron beams was measured by Mott and M$\o$ller scattering. 
Their longitudinal polarization is transfered in the bremsstrahlung process to circular 
polarization of the photons. The polarization degree of the photon beams follows from the 
polarization degree of the electrons and the energy-dependent polarization transfer factors 
given by Olsen and Maximon \cite{Olsen_59}. The beam-helicity asymmetry can then be measured by 
comparing the event rates for the two helicity states of the beam. 
The size of the tagged photon beam spot on the targets was restricted to $\approx$1.3 cm diameter
by a collimator (4 mm diameter) placed downstream from the radiator foil. The targets were Kapton 
cylinders of $\approx$4 cm diameter and different lengths filled with liquid hydrogen or
liquid deuterium. Contributions from the target windows ($2 \times 120\ \mu m$ Kapton) were 
determined with empty target measurements, but are negligible for the results discussed 
in this paper. Data were taken during four different beam times. Their main parameters
are summarized in Table ~\ref{tab:beam}.

\begin{table}[hhh]
\begin{center}
  \caption[Summary of data sets]{
    \label{tab:beam}
     Summary of data samples. Target type ($LD_2$: liquid deuterium, 
     $\rho_d$ = 0.169 g/cm$^3$; $LH_2$: liquid hydrogen, 
     $\rho_H$ = 0.071 g/cm$^3$), target length 
     $\ell$ [cm], target surface density $\rho_s$ [nuclei/barn], 
     electron beam energy $E_{e^-}$ [MeV],
     degree of longitudinal polarization of electron beam $P_{e^-}$ [\%].
}
\vspace*{0.3cm}
\begin{tabular}{|c|c|c|c|c|}
\hline
Target & $\ell$ [cm] & ~$\rho_s$ [barn$^{-1}]~$ & 
$E_{e^-}$ [MeV] & $P_{e^-}$ [\%]\\
\hline\hline
  $LD_2$ & 4.72 & 0.231$\pm$0.005 & 1508 & 61$\pm$4\\
  $LD_2$ & 4.72 & 0.231$\pm$0.005 & 1508 & 84.5$\pm$6\\
  $LD_2$ & 3.00 & 0.147$\pm$0.003 & 1557 & 75.5$\pm$4\\
  $LH_2$ & 10.0 & 0.422$\pm$0.008 & 1557 & 75.5$\pm$4\\  
\hline
\end{tabular}
\end{center}
\end{table}

\begin{figure}[htb]
\resizebox{0.50\textwidth}{!}{%
  \includegraphics{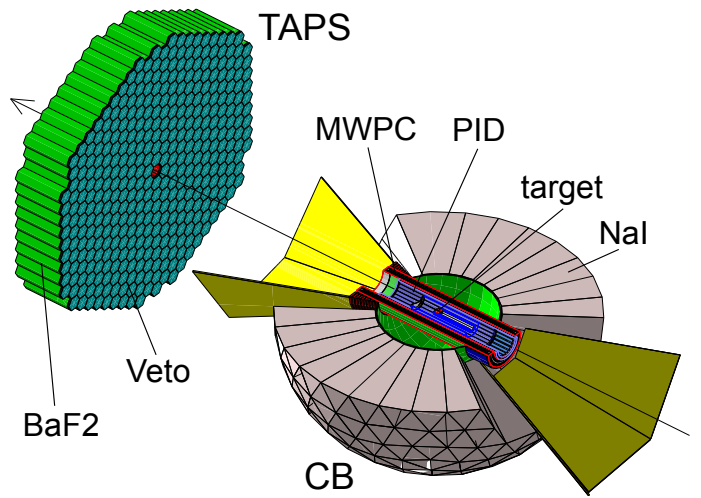}
}
\caption{Experimental setup of Crystal Ball (only bottom hemisphere shown)
with PID detector and TAPS forward wall.
}
\label{fig:setup}       
\end{figure}

Photons, charged pions, and recoil nucleons produced in the target were detected with an
almost $4\pi$ electromagnetic calorimeter schematically shown in Fig.~\ref{fig:setup}.
It combined the Crystal Ball detector (CB) \cite{Starostin_01} with the TAPS detector 
\cite{Novotny_91,Gabler_94}. 
The CB is made of 672 NaI crystals and covers the full azimuthal 
range for polar angles from 20$^{\circ}$ to 160$^{\circ}$, corresponding to 93\% of the full 
solid angle. It is arranged in an upper and lower hemisphere (only the lower hemisphere
is shown in Fig.~\ref{fig:setup}). The TAPS detector, consisting
of 384 BaF$_2$ crystals, was configured as a forward wall, placed 1.457 m downstream from the
targets, and covered polar angles from $\approx~$5$^{\circ}$ to $\approx$~21$^{\circ}$.
The Crystal Ball was equipped with a Particle Identification 
Detector (PID) \cite{Watts_04} for the identification of charged particles and 
all modules of the TAPS detector had individual plastic scintillators in front for the 
same purpose (TAPS `Veto-detector'). This setup is similar to the one described in more detail 
in \cite{Zehr_12,Schumann_10} (the only difference is the size and position of the TAPS 
forward wall) and identical to the setup used for the measurement of the
double-$\pi^0$ final state \cite{Oberle_13}. 

The trigger conditions varied for the four beam times. They were optimized for different reaction 
types ranging from low-multiplicity final states like single $\pi^0$ production, or even Compton 
scattering, to high-multiplicity states like production of $\pi^0$ pairs or 
$\eta\rightarrow 3\pi^0\rightarrow 6\gamma$ decays. They were always based on the multiplicity 
of hits in the combined calorimeter and the analog sum of the energy signals from detector 
modules of the Crystal Ball. For the multiplicity information, both calorimeters were subdivided
into logical sectors. The TAPS detector was divided into 6$\times$64 modules in a pizza-like 
geometry (i.e. into 6 triangularly shaped sectors pointing to the beam pipe) and the CB into 
45 rectangular sectors (each composed of 16 detector modules of triangular cross section). 
The different triggers required hits in 1 - 3 logical sectors 
of the combined calorimeter with analog energy sums in the CB of $\approx$300~MeV. Triggers with 
hit multiplicity of one or two are activated by the decay photons from the $\pi^0\pi^{\pm}$ 
final state. For multiplicity-three triggers the charged pion must also contribute, which resulted 
in larger systematic uncertainties for absolute cross sections. One should, however, keep in mind
that such uncertainties (also from the exact definition of the analog sum threshold of the CB)
cancel in the asymmetries discussed in this paper.  

\section{Data analysis}
\label{sec:ana}

The reactions analyzed were $\gamma p\rightarrow n \pi^0\pi^+$ (photoproduction off free protons), 
$\gamma d \rightarrow (n) n\pi^0\pi^+$ (photoproduction off quasi-free protons bound in the deuteron),
and $\gamma d\rightarrow (p) p\pi^0\pi^-$ (photoproduction off quasi-free neutrons bound in the 
deuteron). The nucleon in brackets is the spectator, the other nucleon the participant in the final state
(the initial-state participant has of course the other charge).
Detection of the participant recoil nucleon is mandatory for reactions measured with the deuteron 
target. Therefore, detection of the recoil neutron was also required for the measurement with the 
hydrogen target so that the analysis for both targets was identical. This means that for all 
reactions, the accepted events were those with candidates for two photons from the $\pi^0$ decay, 
a candidate for a charged pion, and a candidate for either a recoil proton (only for the deuteron 
target) or a recoil neutron. 

\begin{figure}[thb]
\resizebox{0.50\textwidth}{!}{%
  \includegraphics{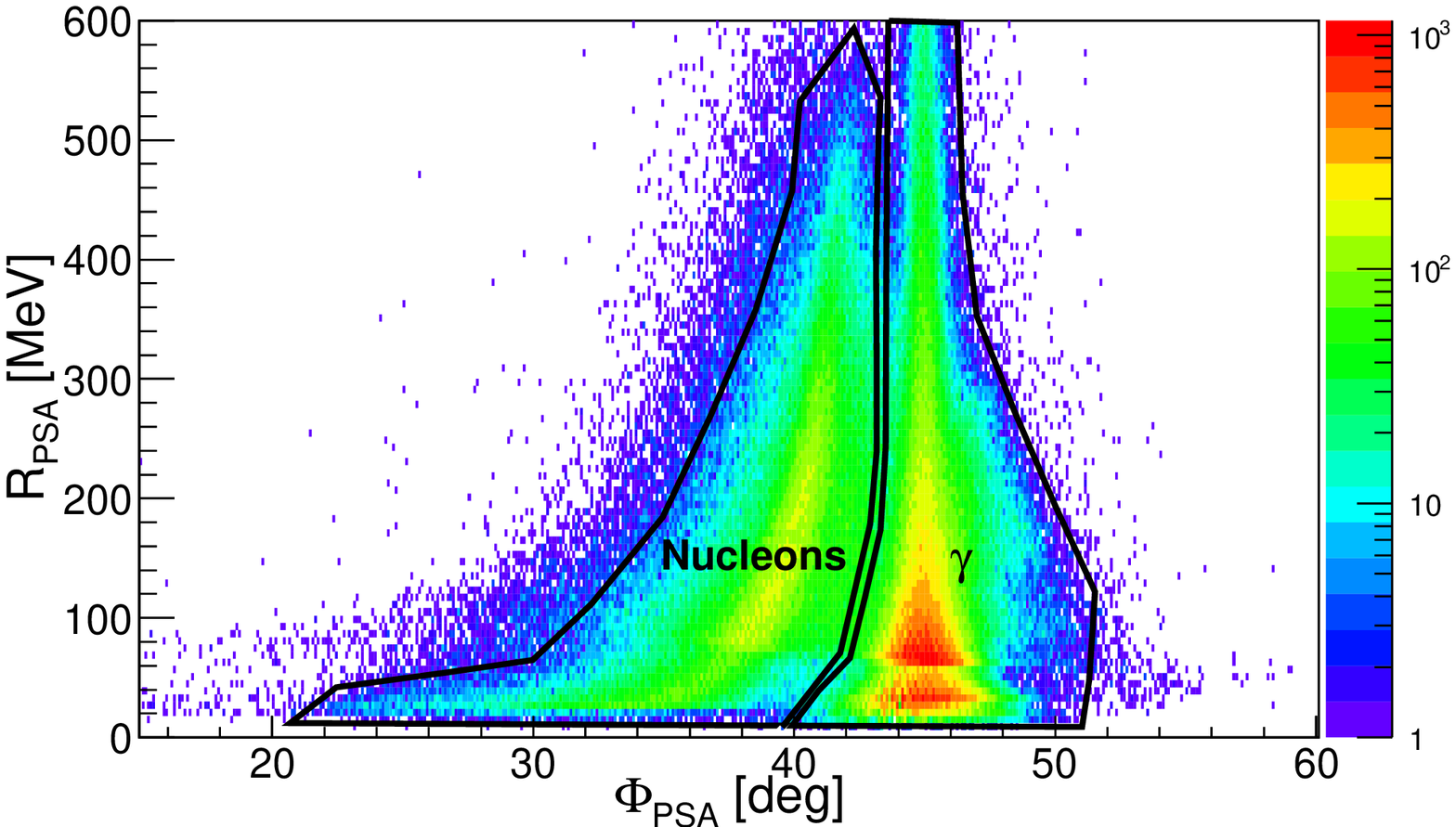}
}
\resizebox{0.50\textwidth}{!}{%
  \includegraphics{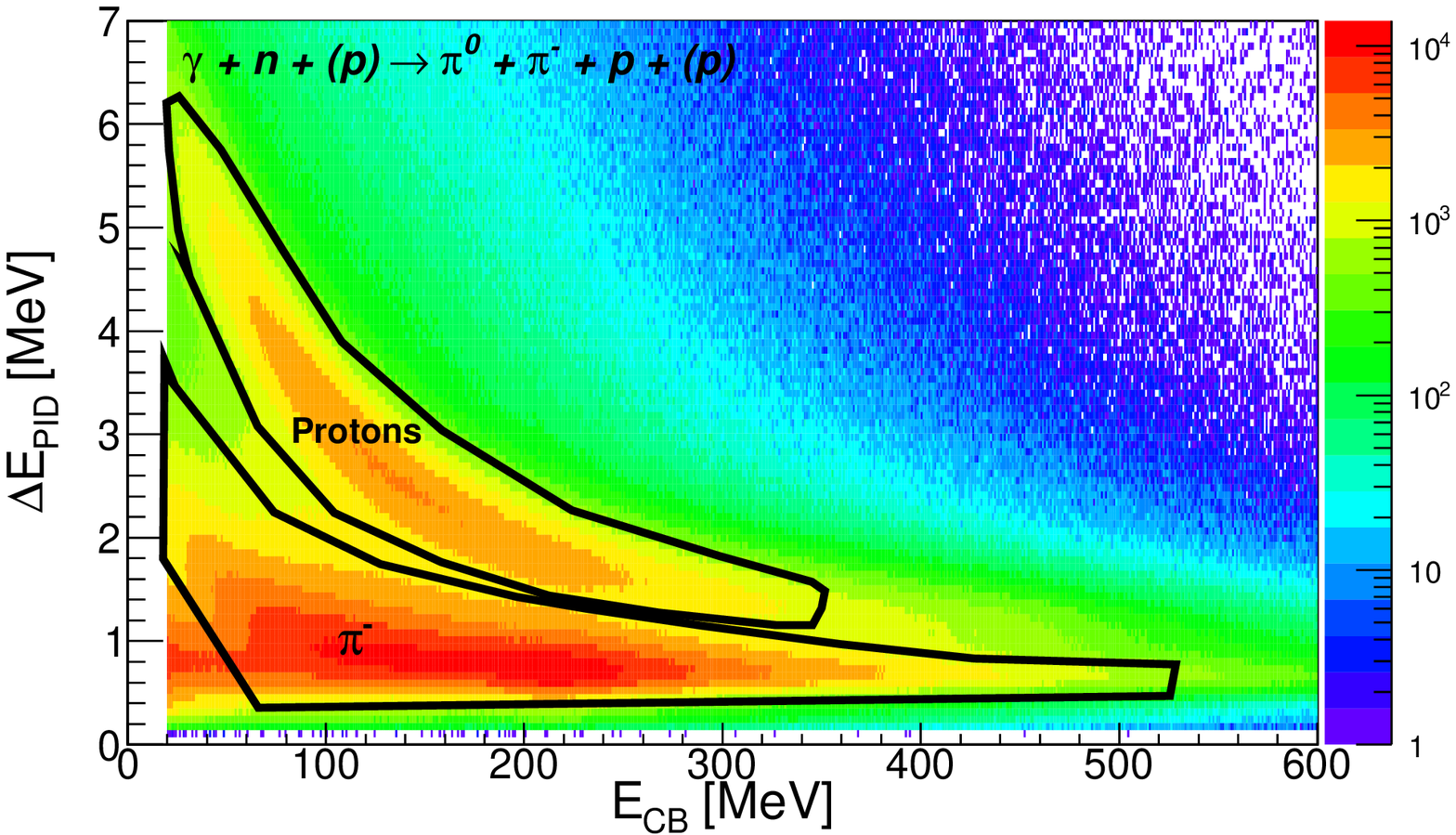}
}
\caption{Upper panel: pulse-shape analysis in TAPS for one individual detector module.
Plotted is the radius $R_{PSA}$ versus the angle $\Phi_{PSA}$ of the polar-coordinate
parameterization of the pulse-shape like in Ref.~\cite{Zehr_12}.
Lower panel: $\Delta E-E$ analysis
with PID and CB. Energy deposited in the PID scintillators as function of the energy deposited 
in the CB. Separation of protons and charged pions for candidates of the $\pi^0\pi^- p$ final 
state (two charged, two neutral hits in the calorimeter). 
}
\label{fig:par_iden}       
\end{figure}

The analysis combined the particle identification possibilities of the detector (charged particle
identification, pulse-shape analysis (PSA) in TAPS, time-of-flight (ToF) versus energy in TAPS, and 
$\Delta E-E$ analysis of CB and PID; see \cite{Zehr_12,Schumann_10} for details) with the reaction 
identification via invariant-mass analyses, meson-pair nucleon coplanarity, and missing-mass analyses.
The separation of photons and recoil nucleons in TAPS via PSA and the separation of recoil protons 
and charged pions in the CB-PID system by the $\Delta E-E$ analysis is shown in Fig.~\ref{fig:par_iden}.
For both reactions, the first step of the analysis used the charged-particle identification sub-detectors 
(TAPS-`Veto' and PID) to assign hits in the ca\-lo\-ri\-meter parts of the detector to `charged' or 
`neutral'. Events with exactly one `charged' and three `neutral' hits were analyzed as candidates for the 
$n\pi^+\pi^0$ final state and events with exactly two `charged' and exactly two `neutral' hits were 
accepted as candidates for the $p\pi^-\pi^0$  final state. 

\begin{figure}[thb]
\resizebox{0.50\textwidth}{!}{%
  \includegraphics{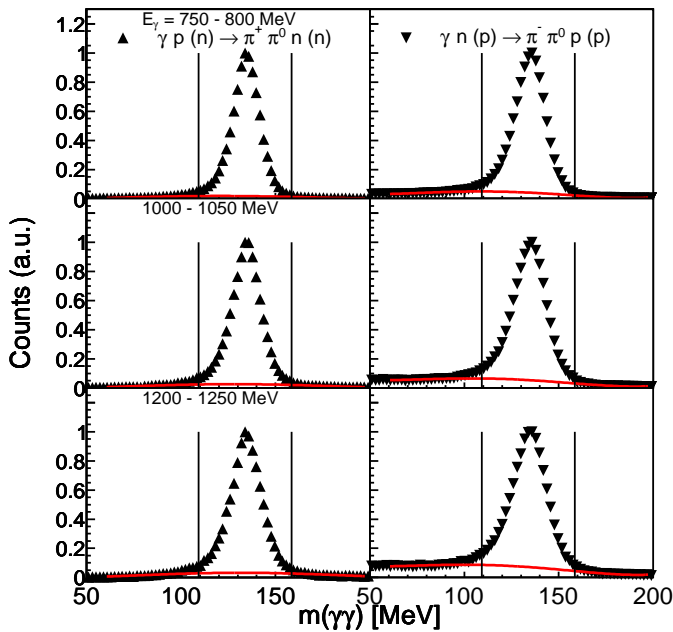}
}
\caption{Invariant-mass distribution of the candidates for two-photon decays of the $\pi^0$.
Left hand side: $\pi^+ \pi^0 n$, right hand side: $p\pi^-\pi^0$. 
For the $\pi^+ \pi^0 n$ final state the `best' $\gamma \gamma$-pair is selected by the $\chi^2$
test. The two vertical lines indicate the accepted range. The red line is a polynomial fit to 
the background. 
}
\label{fig:invm}       
\end{figure}

In the next step, the neutral hits were inspected more closely. For assumed $p\pi^-\pi^0$ events
it was required that the candidates for the $\pi^0$-decay photons, when detected in TAPS, passed the
PSA filter as photons. For candidates of $n\pi^+\pi^0$ with three neutral hits, a $\chi^2$ test
was first used to identify the most probable combination of the three neutral hits to decay photons of a
$\pi^0$ meson and a neutron. This was done by minimizing
\begin{equation}
\chi^2 = \frac{(m_{\gamma\gamma}(k)-m_{\pi^0})^2}{\Delta m_{\gamma\gamma}(k)}\;\; k=1,2,3 \nonumber
\end{equation}
where $m_{\gamma\gamma}(k)$ are the invariant masses of the three possible combinations of neutral hits to 
pion-decay photons, $\Delta m_{\gamma\gamma}(k)$ are their uncertainties, and $m_{\pi^0}$ is the nominal
pion mass. The two neutral hits of the `best' combination were taken as photon candidates, leaving the third hit
as a neutron candidate. Subsequently, for neutral hits in TAPS it was checked whether hits assigned as 
photons passed the PSA analysis cuts for photons and hits assigned to neutrons passed the neutron PSA cut.
For neutral hits in the CB no additional conditions could be applied. The resulting invariant-mass
spectra for both reaction channels are summarized in Fig.~\ref{fig:invm}. The background level is very low.
For both reactions entries with invariant masses between 110~MeV and 160~MeV were accepted. 

The nominal invariant mass $m_{\pi^0}$ of the $\pi^0$ meson was then used to improve the experimental 
resolution further. Since for both sub-calorimeters the angular resolution is better than the energy resolution, 
this was simply done by replacing the measured energies $E_{i}$ of the photon hits by
\begin{equation}
E^{'}_{i}\ =\ E_{i}\frac{m_{\pi^{0}}}{m_{\gamma\gamma}}\;\; i=1,2  \nonumber
\end{equation}
where $m_{\gamma \gamma}$ are the measured invariant masses.

Subsequently, the candidates for protons and charged pions were analyzed. The separation of protons and 
charged pions in CB with help of the CB-PID $\Delta E - E$ analysis was very efficient, but the separation in 
TAPS via ToF versus energy was not as good. Due to the high intensity in the proton band (partly from 
background reactions) the pion band in ToF versus energy was contaminated with protons. Therefore, events 
with the charged pion candidate in TAPS were not included in the analysis. The result is that a small part 
of the reaction phase-space (polar laboratory angles of charged pion $<$ 20$^{\circ}$) was excluded. 
This is only a small effect, but must be taken into account when the results are compared to model 
predictions. Events accepted for $n\pi^+\pi^0$ were those with the charged pion identified in the CB via 
$\Delta E-E$. For $p\pi^-\pi^0$ events, it was required that the charged pion satisfied the $\Delta E-E$ 
condition and that the other charged hit (proton candidate) when detected in the CB passed 
the $\Delta E-E$ analysis as proton, or, when in TAPS, passed the PSA filter as nucleon. 

\begin{figure*}[thb]
\centerline{\resizebox{1.0\textwidth}{!}{%
  \includegraphics{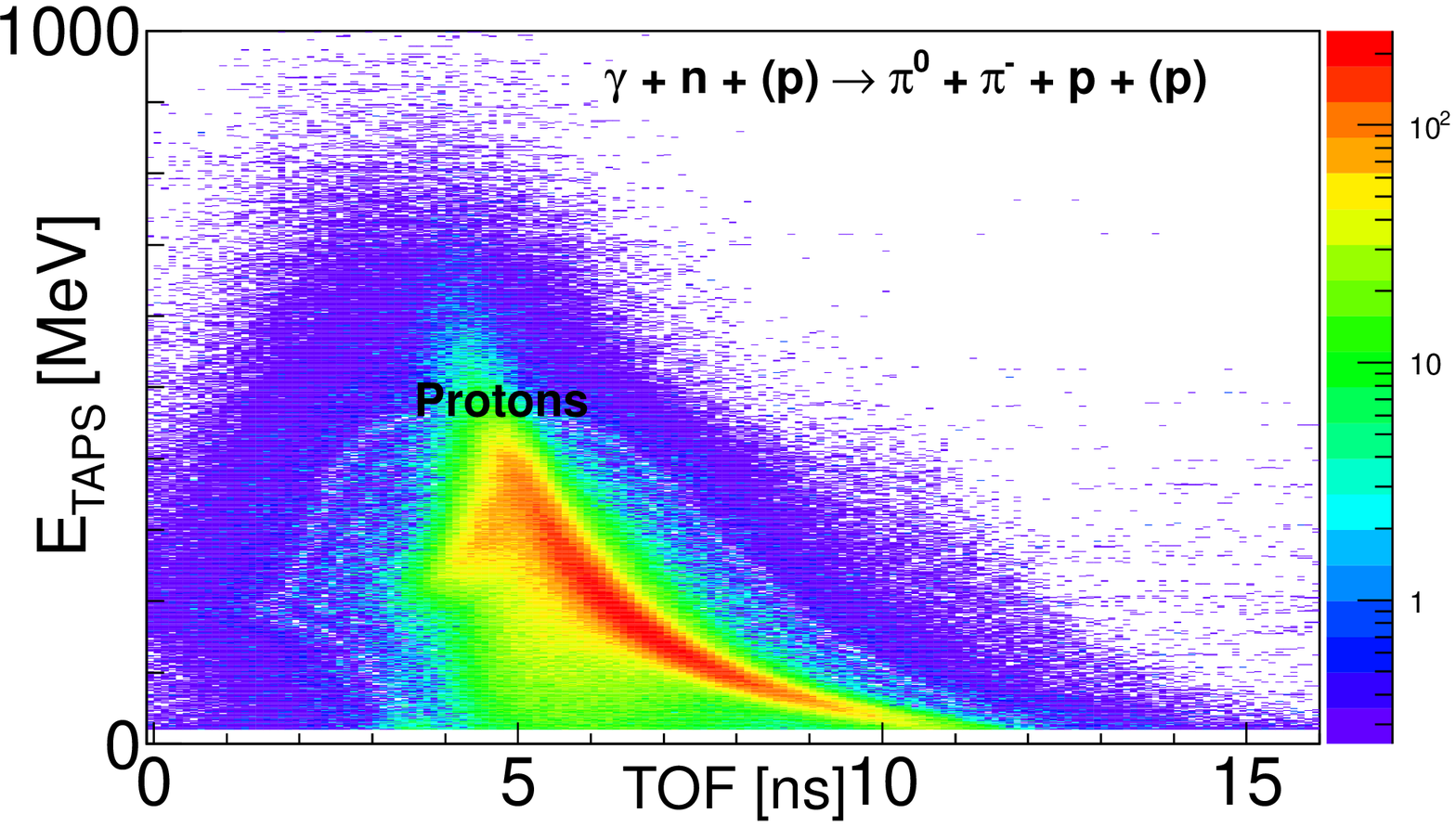}
  \includegraphics{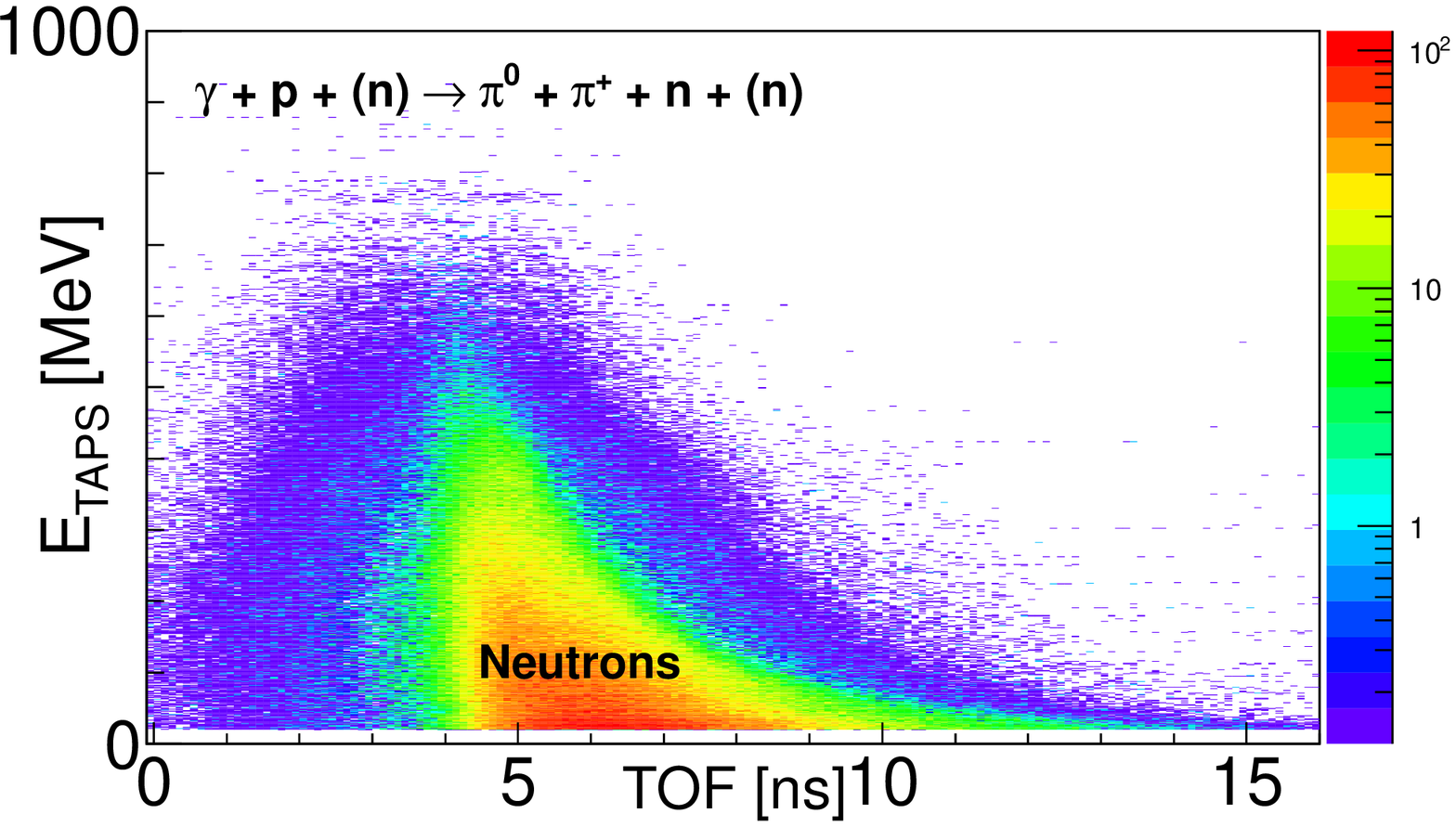}  
  \includegraphics{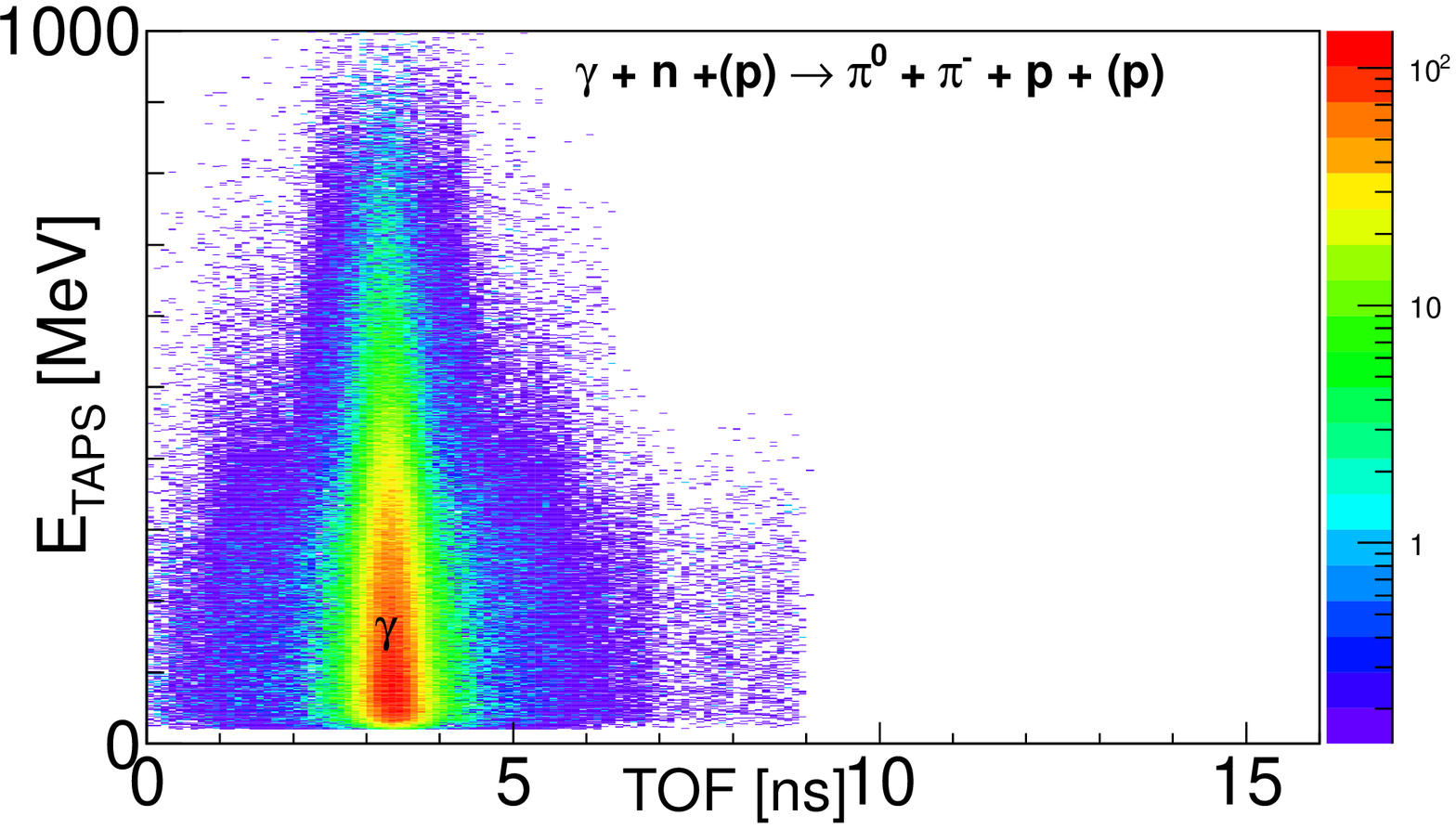}
}}  
\caption{ToF-versus-energy spectra for hits in TAPS assigned to (from left to right) proton, neutron
and photon candidates. 
}
\label{fig:ToF}       
\end{figure*}

For events detected in TAPS, the ToF-versus-energy spectra served as a final test for the particle
identification. Such spectra are summarized in Fig.~\ref{fig:ToF} and show the expected behavior:
photon candidates form a band at constant ToF corresponding to the (normalized) target - detector distance.
Protons are lying in a band matching the relativistic ToF-energy relation for kinetic energies
below $\approx$ 400~MeV. For higher kinetic energies, the band bends back because the protons are no longer 
stopped in the BaF$_2$ crystals but punch through the backside of the detector. Neutrons deposit a random
fraction of their kinetic energy and thus appear in the region below the proton band. Since no
significant background structures were observed in these spectra, no cuts were applied in order to
avoid unnecessary systematic uncertainties related to the cuts. Cutting roughly on the signal regions
in the spectra has no effect on the results.

Altogether, at this stage of analysis, the identification of the different particle types with partly 
redundant filters is excellent. However, there is still background from competing reactions where some
final-state particles escaped detection. These are, for example, events from triple-pion
production (a significant fraction stems from the $\eta\rightarrow\pi^0\pi^+\pi^-$ decay), where
one charged pion was too low in energy for detection or went along the beam-pipe. 

\begin{figure}[thb]
\resizebox{0.48\textwidth}{!}{%
  \includegraphics{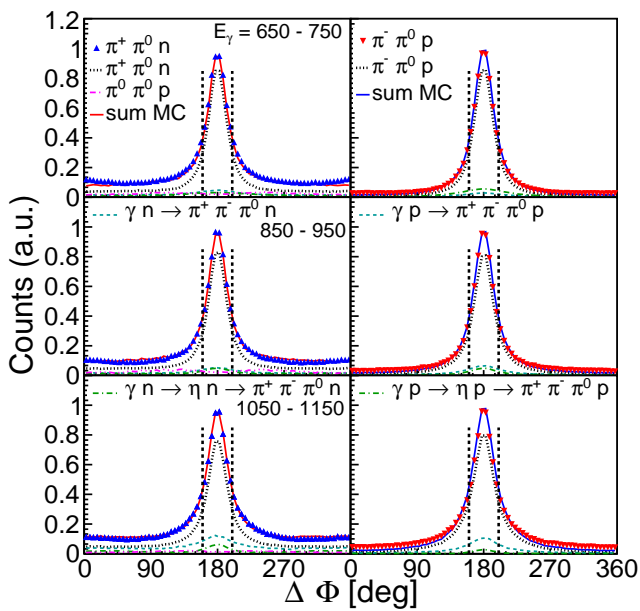}
}
\caption{Spectra of the azimuthal difference between the cm three-momenta of the two-pion system and the
recoil nucleon. Left panel: recoil neutrons, right panel: recoil protons. Triangles: measured data,
curves: MC simulations of signal and background components. 
}
\label{fig:copla}       
\end{figure}

Such background must be eliminated using the overdetermined reaction kinematics. As a first step, the
coplanarity of the meson pairs with the recoil nucleon was analyzed. The sum of the three-momentum 
components of the $N\pi\pi$ final-state particles perpendicular to the beam axis must vanish
(apart from effects from Fermi motion, which broaden the distributions). Figure~\ref{fig:copla} shows
the azimuthal angular difference $\Delta\Phi$ of the three-momenta of the two-pion system and the recoil 
nucleon in the cm system together with Monte Carlo (MC) simulations of the expected signal shape and background 
contributions. Only events with $\Delta\Phi$ in the range (180$\pm$20)$^{\circ}$ were accepted.
This cut removes mainly background for the reaction with coincident recoil neutrons, but cannot 
completely suppress it since some background contributions such as events from
$\eta \rightarrow 3\pi$ also peak at 180$^{\circ}$. For the $\pi^-\pi^0 p$ final state this is the 
dominant background contribution (mainly arising from the final state $n\pi^0\pi^+\pi^-$ when the 
recoil neutron escapes detection and one of the charged pions is misidentified as a proton). 
However, this background is subsequently removed by the more efficient missing-mass analysis 
(see below).
 
\begin{figure}[thb]
\resizebox{0.48\textwidth}{!}{%
  \includegraphics{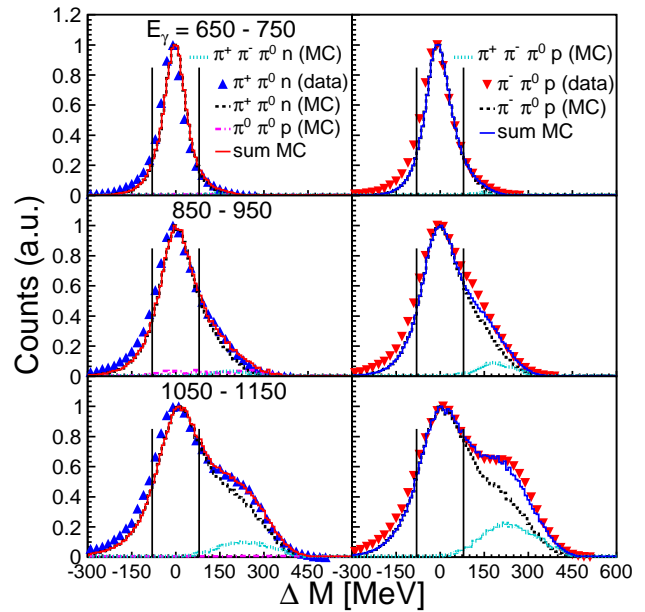}
}
\caption{Missing-mass distribution for three typical ranges of $E_{\gamma}$. 
Left column: (blue) triangles experimental results for $\pi^0\pi^+n$ final state. 
Dotted (black) lines: MC signal,  (light blue) lines: MC background from triple pion production,
(magenta) lines: MC background from $\pi^0\pi^0p$, solid  (red) lines: sum of MC signal 
and MC backgrounds, vertical (black) lines: applied cut. Right column: (red) triangles: experimental
results for $\pi^0\pi^-p$ final state. Solid (blue) lines: sum of MC signal and backgrounds. 
}
\label{fig:mism}       
\end{figure}

In the final analysis step the recoil nucleons, although detected, were treated as missing particles and 
their mass $m_N$ was compared to the mass reconstructed from the incident photon energy and the 
four-vectors of the two pions via
\begin{equation}
 \Delta M =\ \left | P_{\gamma}\ +\ P_N\ -\ P_{\pi^0}\ -\ P_{\pi^{\pm}}\right | - m_N
\end{equation}
where $P_\gamma$, $P_{N}$ are the four-vectors of incident photon and incident nucleon (assumed to be 
at rest, with the distribution again broadened by Fermi motion), and $P_{\pi^0}$, P$_{\pi^{\pm}}$ are the 
four-momenta of the pions.  
The result of this analysis is summarized in Fig.~\ref{fig:mism} and compared to simulations of the 
signal shape and background from triple-pion production either from the $\eta$-decay or from phase-space 
contributions which are the main background sources. In case of the $\pi^0\pi^+ n$ final state there is 
also a small background component from the $\gamma p\rightarrow\pi^0\pi^0p$ reaction with one undetected 
photon, the proton misidentified as charged pion, and one photon misidentified as a neutron. The sum of 
the MC simulations for signal and background does a good job of reproducing the measured data. 
For the construction of the asymmetries events were only accepted in the $\Delta M$ range 
($-80$ MeV to $+80$ MeV) for which the simulations indicated very small background. The cut at -80 MeV 
does not improve the peak-to-background ratio but avoids systematic effects from Fermi motion. With an 
asymmetric cut, one would select a biased momentum distribution of the nucleons (preferring nucleon 
momenta antiparallel to the photon momentum). Since the statistical quality of the data is excellent, 
the small loss in counting statistics did not matter.    

\begin{figure}[thb]
\centerline{
\resizebox{0.49\textwidth}{!}{%
  \includegraphics{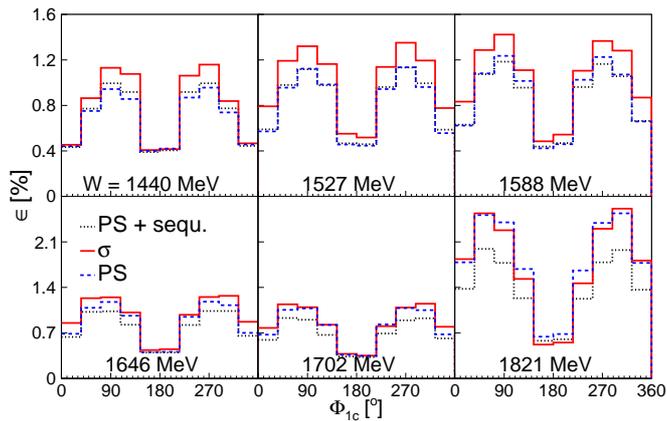}
}}
\caption{Simulated detection efficiency as function of the angle $\Phi_{1c}$ for different bins of
final state invariant mass for the reaction $\gamma p\rightarrow n\pi^0\pi^+$ for the free proton target.
Solid (red) histograms: event generator from model \cite{Fix_05},
dashed (blue): phase-space, dotted (black): phase space and sequential decays via $\Delta$(1232)
intermediate state.   
}
\label{fig:eff_w}       
\end{figure}

\begin{figure}[htb]
\centerline{
 \resizebox{0.47\textwidth}{!}{%
  \includegraphics{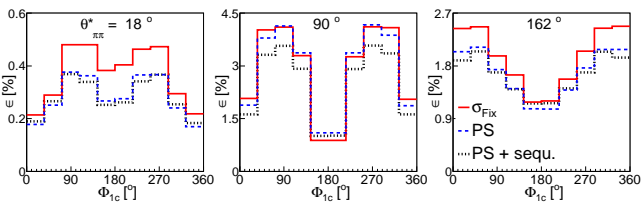}
}}  
\caption{Same as Fig.~\ref{fig:eff_w} but for different bins of the polar angle of the parent particle
of the two pions (i.e. 180$^{\circ}$ - $\Theta^{\star}_N$). 
}
\label{fig:eff_t}       
\end{figure}

In order to remove completely the influence of nuclear Fermi motion, the invariant mass $W$ of the 
$N_{p}\pi\pi$ final state ($N_{p}$: participant nucleon) for quasi-free production off the deu\-ter\-on 
was derived event-by-event from the four-momenta of the three particles.
The three momenta of the pions were directly measured with the calorimeter. Azimuthal and polar angles
for all recoil nucleons were measured with good resolution. In principle, kinetic energies
of recoil nucleons detected in TAPS can be reconstructed from ToF. Kinetic energies of protons
up to $\approx$~400~MeV (at higher energies they are not stopped) can be extracted from their
deposited energies. However, for the recoil neutrons registered in the CB only the angles were available. 
Therefore, in order to minimize systematic effects, all quasi-free recoil nucleons were treated in the 
same way and only the measured polar and azimuthal angles were used in the analysis. The kinetic 
energies were then reconstructed from energy and momentum conservation as discussed in 
\cite{Krusche_11,Jaegle_11}. For the measurement with free protons, $W$ was calculated from 
the incident photon energy. 

Asymmetries for a narrowly restricted range of kinematic variables can be constructed from the 
measured count rates according to Eq.~\ref{eq:circ} because all normalization factors such as photon flux,
target density, and detection efficiency cancel in the ratio. However, variations of the detection efficiency
can matter for asymmetries integrated over angles and/or incident photon energies. 
Particularly, the detection efficiency of recoil nucleons varies systematically with their kinetic 
energies and thus also with their polar angles. Therefore, the detection efficiency was simulated with 
the Geant4 code \cite{Geant4}, taking into account all details of the setup. The measured data and also 
the simulated events were analyzed in bins of the final state invariant mass $W$, the angle $\Phi$
between the two planes (see Fig.~\ref{fig:def}), and the cm polar angle of the two-pion system
$\Theta^{\star}_{\pi\pi}=180^{\circ} - \Theta^{\star}_N$, where $\Theta^{\star}_N$ is the cm polar angle 
of the recoil nucleon.
The measured count rates for the three-dimensional cells were then corrected by the simulated detection 
efficiencies for the same cells, projected onto the $\Theta^{\star}$ axis, and into the bins of $W$ 
specified in the figures. Subsequently, the integrated asymmetries were calculated with Eq.~\ref{eq:circ}. 
Since photoproduction of pion pairs involves five independent kinematic variables \cite{Roberts_05},
and the detection efficiency was corrected only in a three-dimensional space (spanned by the three most
important variables), the result depends in principle on the event generator used for the MC simulations. 
Three different event generators were tested. The most simple one used a phase-space distribution of events. 
The second one used a mixture of phase space and the reaction chains 
$\gamma N\rightarrow \pi^0\Delta(1232)\rightarrow \pi^0\pi^{\pm} N$ and
$\gamma N\rightarrow \pi^{\pm}\Delta(1232)\rightarrow \pi^{\pm}\pi^0 N$, where the relative size of the 
contributions from the three processes were fixed by fits of the pion-pion and pion-nucleon invariant 
mass distributions. The third one used the distributions from the Two-Pion-MAID model \cite{Fix_05}. 

\begin{figure}[thb]
\centerline{
\resizebox{0.49\textwidth}{!}{%
  \includegraphics{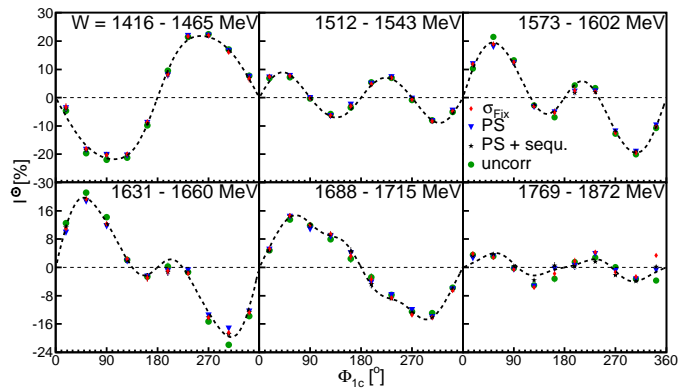}    
}}
\caption{Asymmetry $I^{\odot}_{1c}(\Phi_{1c})$ for $\gamma p\rightarrow n\pi^0\pi^+$ for the free proton 
target and with different detection efficiency corrections.
Color code like in Figs.~\ref{fig:eff_w},\ref{fig:eff_t} and additionally (filled, green circles)
without any efficiency correction. Dotted lines: fits to data (black stars).
All uncertainties only statistical.
}
\label{fig:eff_s}       
\end{figure}

Typical examples of the simulated efficiencies for the reaction $\gamma p\rightarrow n\pi^0\pi^+$
are shown in Fig.~\ref{fig:eff_w} for bins of the final state invariant mass $W$,
and in Fig.~\ref{fig:eff_t} for bins of the cm polar angle $\Theta^{\star}_{\pi\pi}$    
of the two-pion system, both as function of the angle $\Phi_{1c}$ (all other kinematic parameters
integrated out). The efficiencies generated with the different inputs differ in absolute magnitude. 
However, the magnitude of the detection efficiencies and their variation with $\Phi$ does not matter 
here because they cancel in the ratio (see Eq. \ref{eq:circ}). Only their variation with other 
kinematic parameters, which have been integrated out, could matter when the asymmetry changes 
significantly with them. But these effects turn out to be small. As an example, the results for 
$I^{\odot}_{1c}(\Phi_{1c})$ for $\gamma p\rightarrow n\pi^0\pi^+$ for the free proton 
target extracted with the different detection efficiencies are compared in Fig.~\ref{fig:eff_s}.
Also shown in the figure are asymmetries extracted without {\it any} correction for detection efficiency. 
The results are very similar, demonstrating that the efficiency corrections are not critical. 
The main effects from the detection efficiency cancel in the ratio; even for the angle integrated 
count rates. Effects from detection efficiency are also small for the other reactions and asymmetries. 
All results discussed below have been obtained with an efficiency correction using the phase-space event
generator. 

\begin{figure}[thb]
\centerline{
\resizebox{0.49\textwidth}{!}{%
  \includegraphics{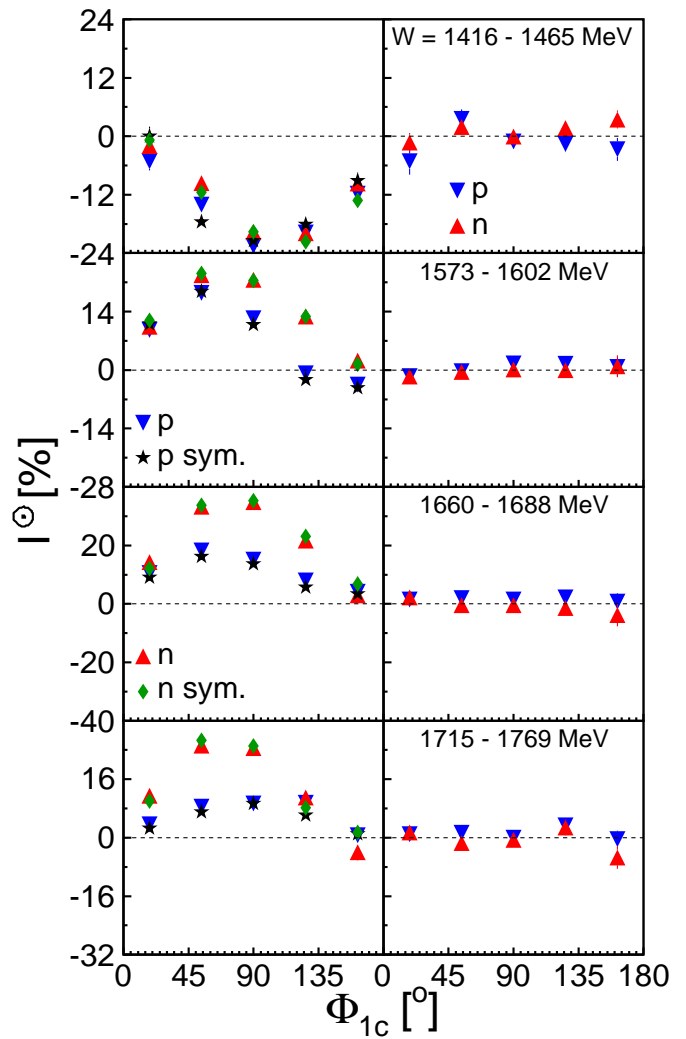}    
}}
\caption{Parity conservation for the asymmetries  $I^{\odot}_{1c}(\Phi_{1c})$. 
Left hand side: $I^{\odot}_{1c}(\Phi_{1c})$, $0^{\circ}\leq\Phi_{1c}\leq 180^{\circ}$ for  
$\gamma p\rightarrow n\pi^0\pi^+$ (blue, down pointing triangles) and for 
$\gamma n\rightarrow p\pi^0\pi^+$ (red, up pointing triangles) compared to 
$-I^{\odot}_{1c}(360^{\circ}-\Phi_{1c})$ (black stars for $\gamma p\rightarrow n\pi^0\pi^+$, 
green diamonds for $\gamma n\rightarrow p\pi^0\pi^+$ . Right hand side: 
$I^{\odot}_{1c}(\Phi_{1c})+I^{\odot}_{1c}(360^{\circ}-\Phi_{1c})$, 
$0^{\circ}\leq\Phi_{1c}\leq 180^{\circ}$ for $\gamma p\rightarrow n\pi^0\pi^+$ 
(blue, down pointing triangles) and $\gamma n\rightarrow p\pi^0\pi^+$ (red, up pointing triangles). 
}
\label{fig:parity}       
\end{figure}

In the following section, only statistical uncertainties are plotted for all results.
The use of Eq.~\ref{eq:circ} assumes of course that the incident photon flux is equal for both 
polarization states of the beam. The polarization state was switched in a randomized way with a 
frequency of 1 Hz. Possible differences in the numbers of incident photons for the two helicity 
states have been determined to be at the 5$\times$10$^{-4}$ level, i.e. they are negligible here. 
The polarization degree of the electron beam was measured with uncertainties between 5\% and 7\%, 
uncertainties arising from the above efficiency correction are estimated below the 5\% level, 
and possible residual background contributions at maximum $W$ are estimated at the 5\% level
(they are negligible for the lowest $W$ values).

\section{Results}

Before we summarize and discuss the extracted asymmetries some remarks to their internal 
consistency and a comparison to the existing data base are appropriate. 

As mentioned in Section \ref{sec:I_intro} parity conservation requires that all asymmetries 
respect Eq.~\ref{eq:sym1}. This condition can be used as an independent test of systematic
uncertainties. All data sets respect this relation within experimental uncertainties, most
already within statistical uncertainties. 
As an example we show in Fig.~\ref{fig:parity} for a few energy bins the asymmetry
$I^{\odot}_{1c}(\Phi_{1c})$ for quasi-free protons and quasi-free neutrons. At the left hand side 
of the figure $I^{\odot}_{1c}(\Phi_{1c})$ is compared to the mirrored values 
$-I^{\odot}_{1c}(360^{\circ}-\Phi_{1c})$ and at the right hand side the sum 
$I^{\odot}_{1c}(\Phi_{1c})+I^{\odot}_{1c}(360^{\circ}-\Phi_{1c})$ is shown. 
The magnitude of the asymmetry is substantial and the sum of original and mirrored values is 
consistent with zero. No systematic discrepancies between the $\Phi =0^{\circ} - 180^{\circ}$
and the $\Phi =180^{\circ} - 360^{\circ}$ data were observed and thus no indication for
false, detector related asymmetries was found. 

\begin{figure}[thb]
\centerline{
\resizebox{0.49\textwidth}{!}{%
  \includegraphics{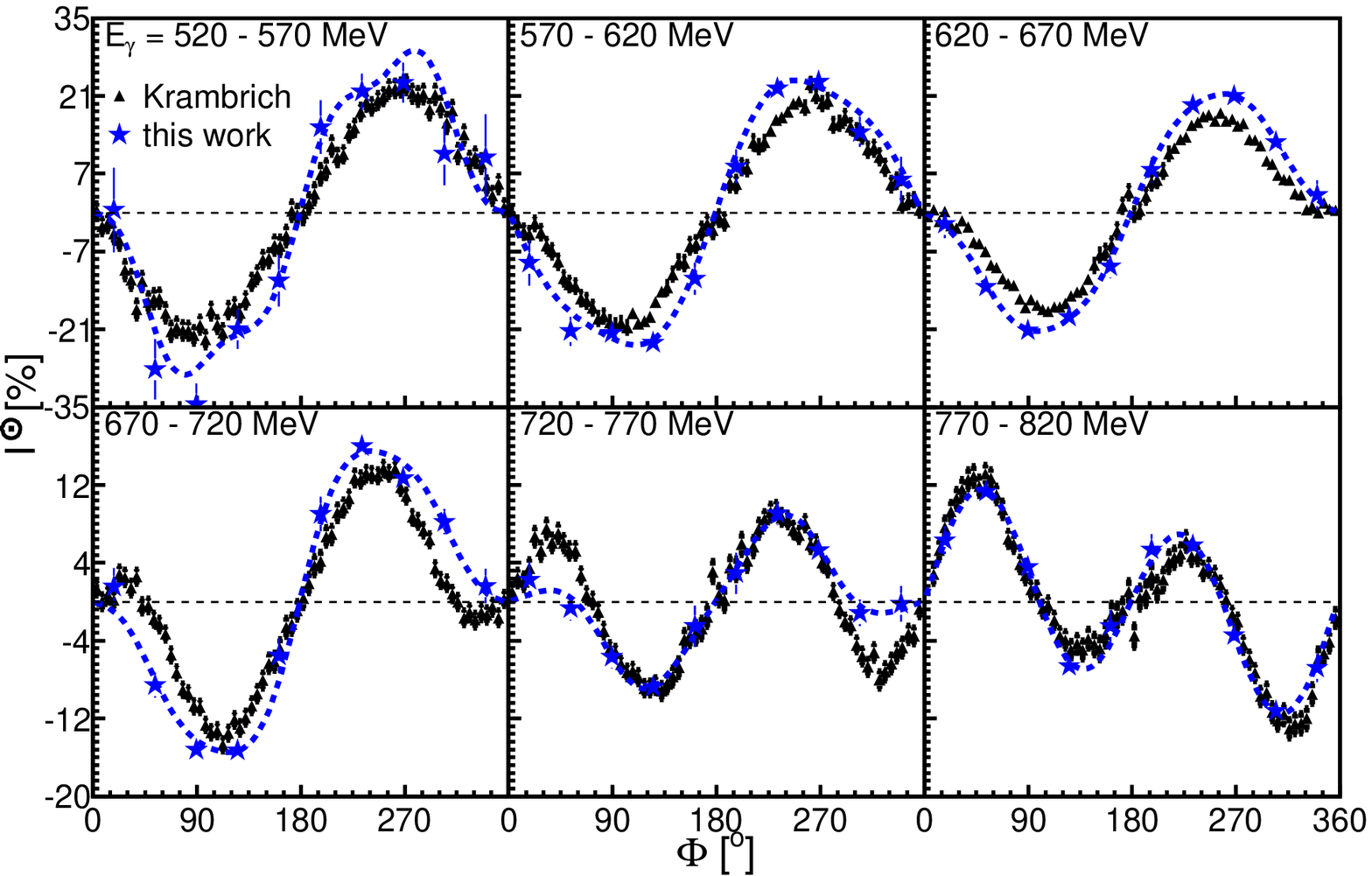}
}}
\caption{Asymmetries (`charge-ordered') for the free proton (blue stars, present experiment)
compared to previous results (black triangles) \cite{Krambrich_09}.
Dashed curve: fits to the data with Eq.~\ref{eq:coeff}.  
}
\label{fig:kra}       
\end{figure}

Previous results are only available for the asymmetry $I^{\odot}_{1c}(\Phi_{1c})$ for the free
proton target and incident photon energies below 820 MeV \cite{Krambrich_09}. They are compared 
in Fig.~\ref{fig:kra} to the present data. The two data sets are in reasonable agreement, but 
the previous data have much better statistical quality. Small systematic discrepancies might arise 
from the different analysis strategies: unlike in the present analysis, in Ref. \cite{Krambrich_09} 
detection of the recoil proton was not required, which removes one source of possible systematic
effects. Detection efficiency effects were not considered in \cite{Krambrich_09}, but as discussed
above they seem to be negligible. One should also note that the lowest energy bins shown in this 
figure are at the very limit accessible by the present experiment (mainly due to the trigger 
conditions which required an energy deposition of 300~MeV in the Crystal Ball), while the previous 
experiment was optimized for the low energy range.

\begin{figure*}[!t]
\centerline{\resizebox{0.99\textwidth}{!}{%
  \includegraphics{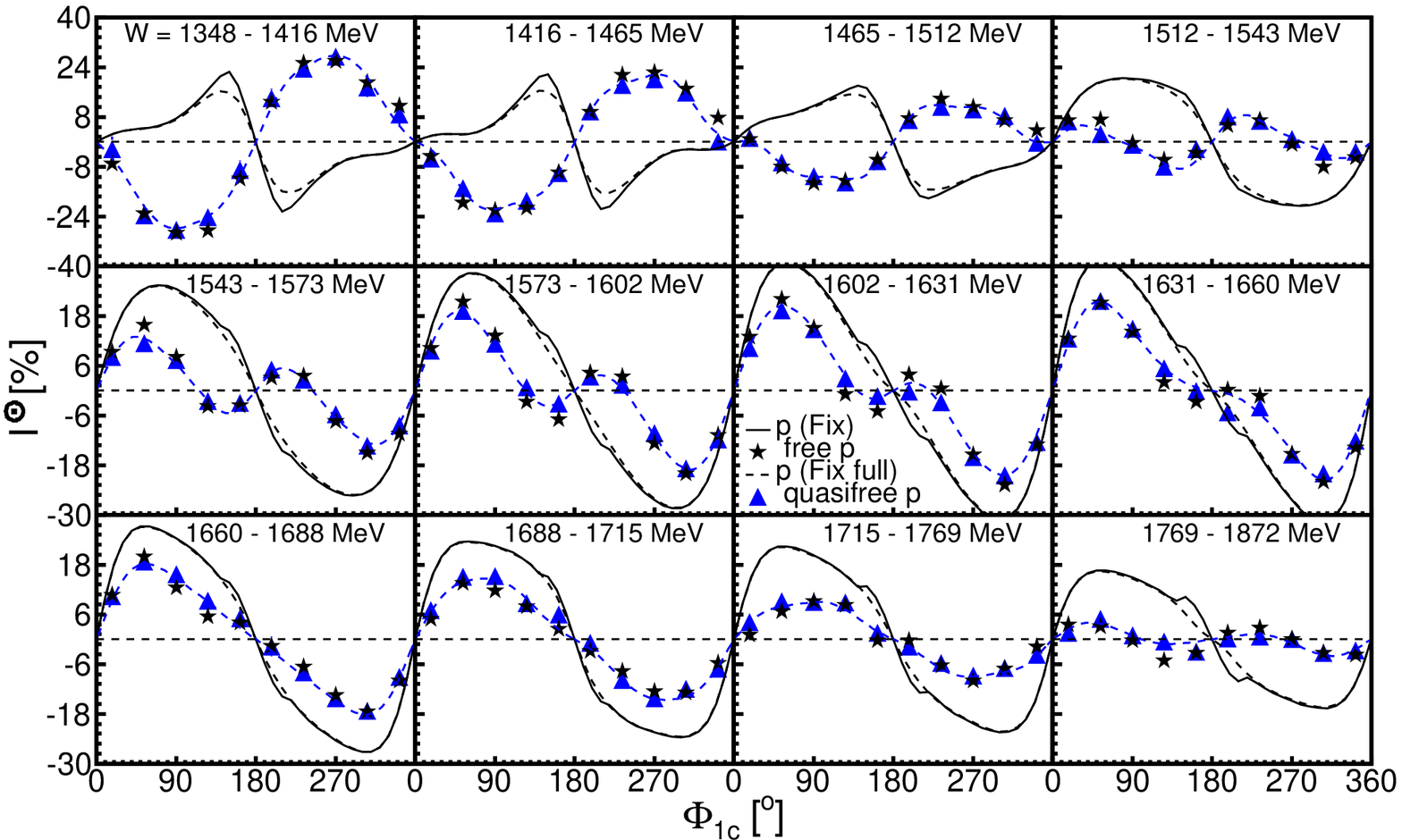}
}}
\centerline  {\resizebox{0.99\textwidth}{!}{%
\includegraphics{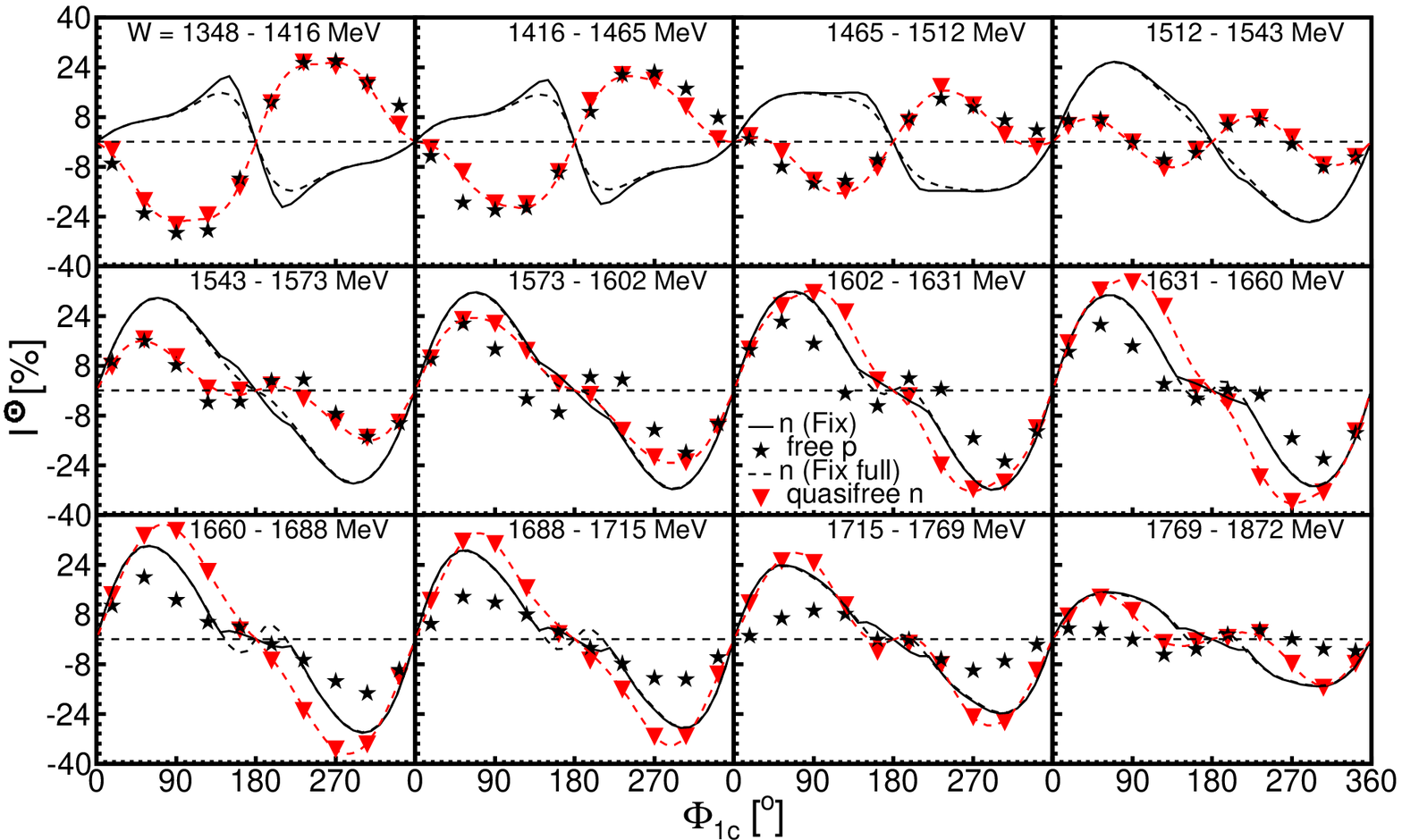}
}}
\caption{Results for `charge-ordered' $I^{\odot}_{1c}(\Phi_{1c})$ for different ranges 
of $W=\sqrt{s}$. 
Upper block: (black) stars: free proton, (blue) triangles: quasi-free proton.
Dashed (blue) curves: fits to quasi-free proton data with Eq.~\ref{eq:coeff}.
(Black) solid curves: model results from \cite{Fix_05} taking into account
experimental acceptance. (Black) dashed: model results without acceptance
restriction. 
Lower Block: data for quasi-free neutrons (red) triangles compared to free proton.
Dashed (red) curves: fits to neutron data. Solid, dashed (black) curves: model 
results from \cite{Fix_05}. 
}
\label{fig:integ_asym_pi}       
\end{figure*}

\begin{figure*}[!t]
\centerline{\resizebox{0.99\textwidth}{!}{%
  \includegraphics{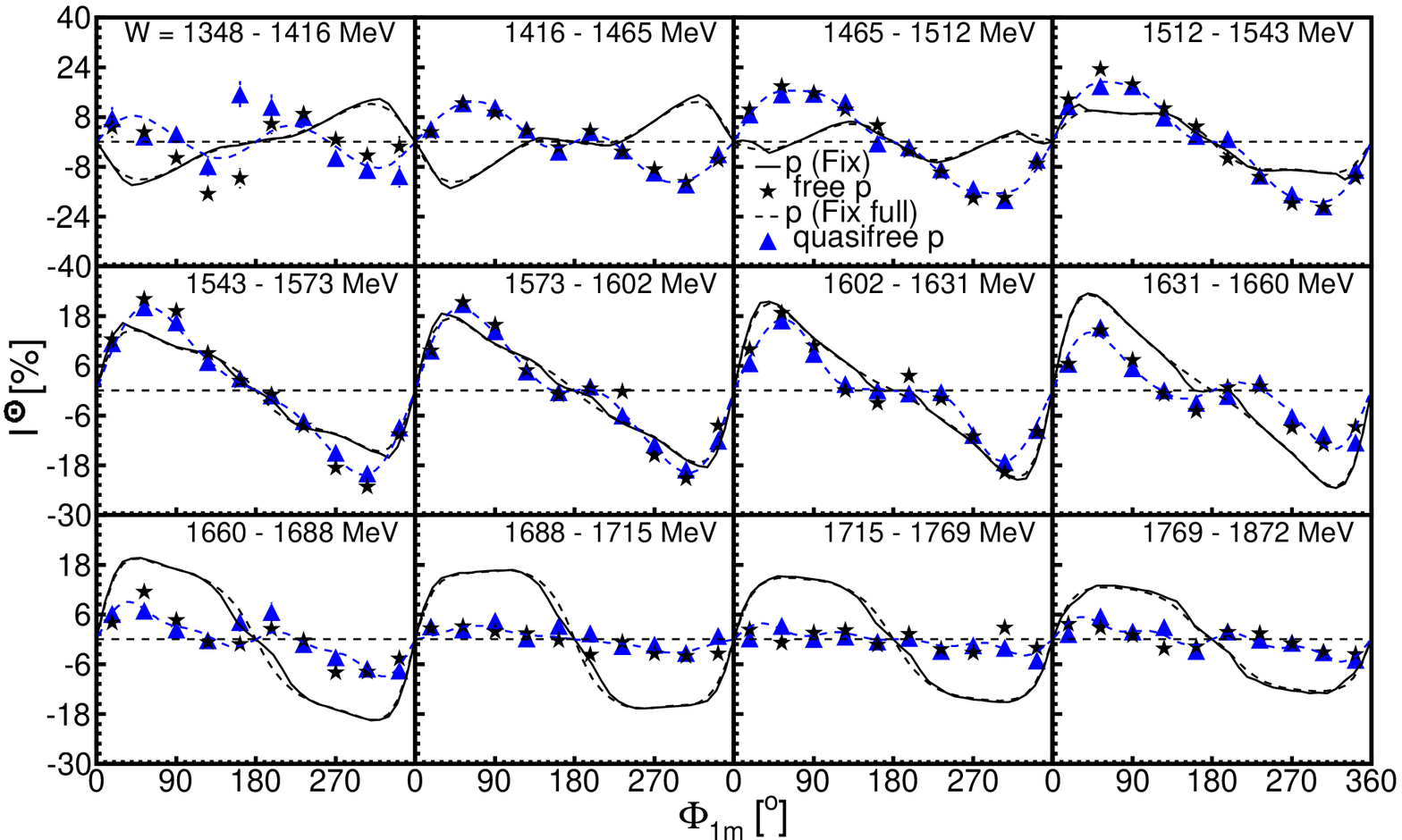}
}}
\centerline  {\resizebox{0.99\textwidth}{!}{%
\includegraphics{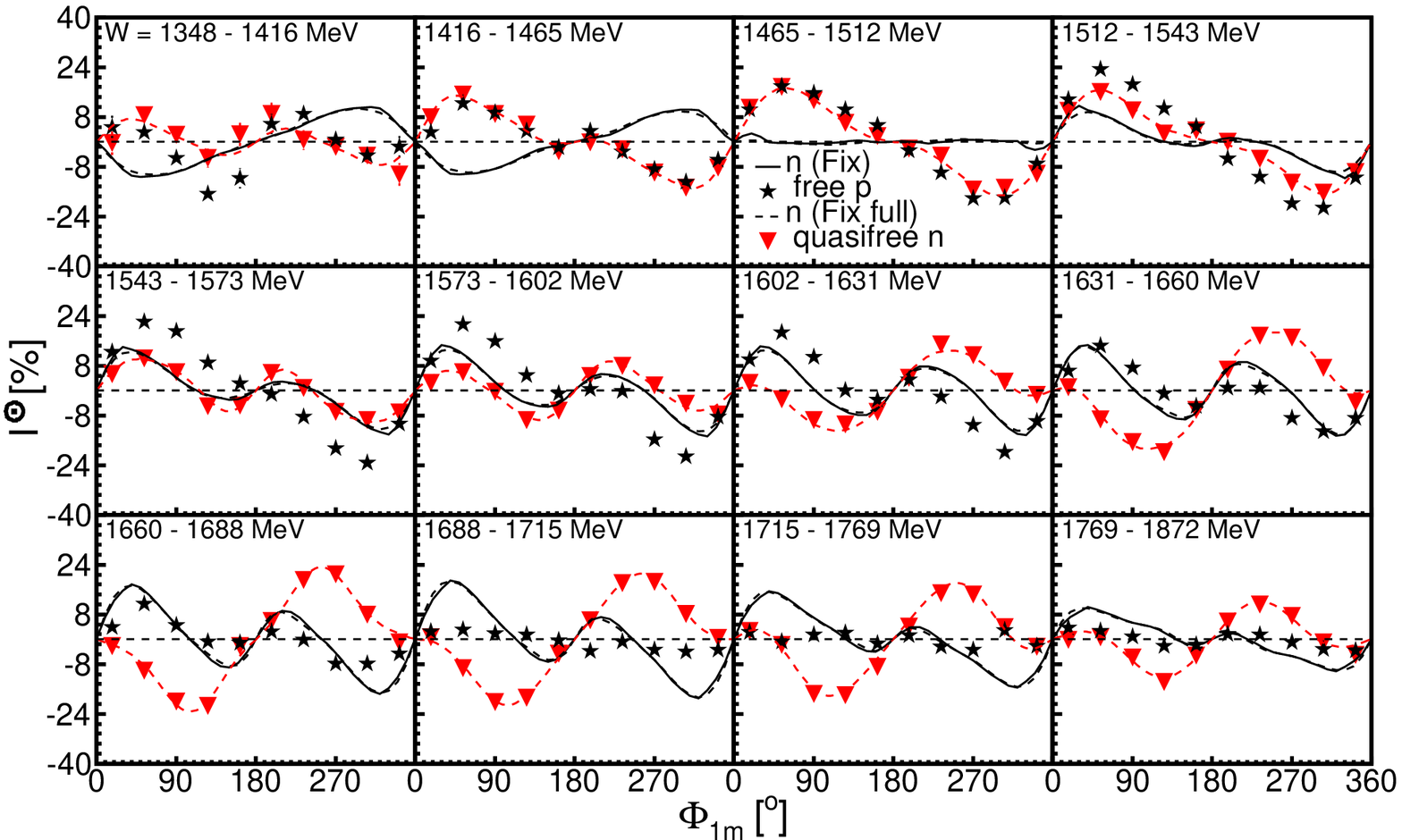}
}}
\caption{Results for `mass-ordered' $I^{\odot}_{1m}(\Phi_{1m})$ for different ranges of $W=\sqrt{s}$. 
For labeling see Fig.~\ref{fig:integ_asym_pi}. 
}
\label{fig:integ_asym_im}       
\end{figure*} 

\begin{figure*}[!t]
\centerline{\resizebox{0.99\textwidth}{!}{%
  \includegraphics{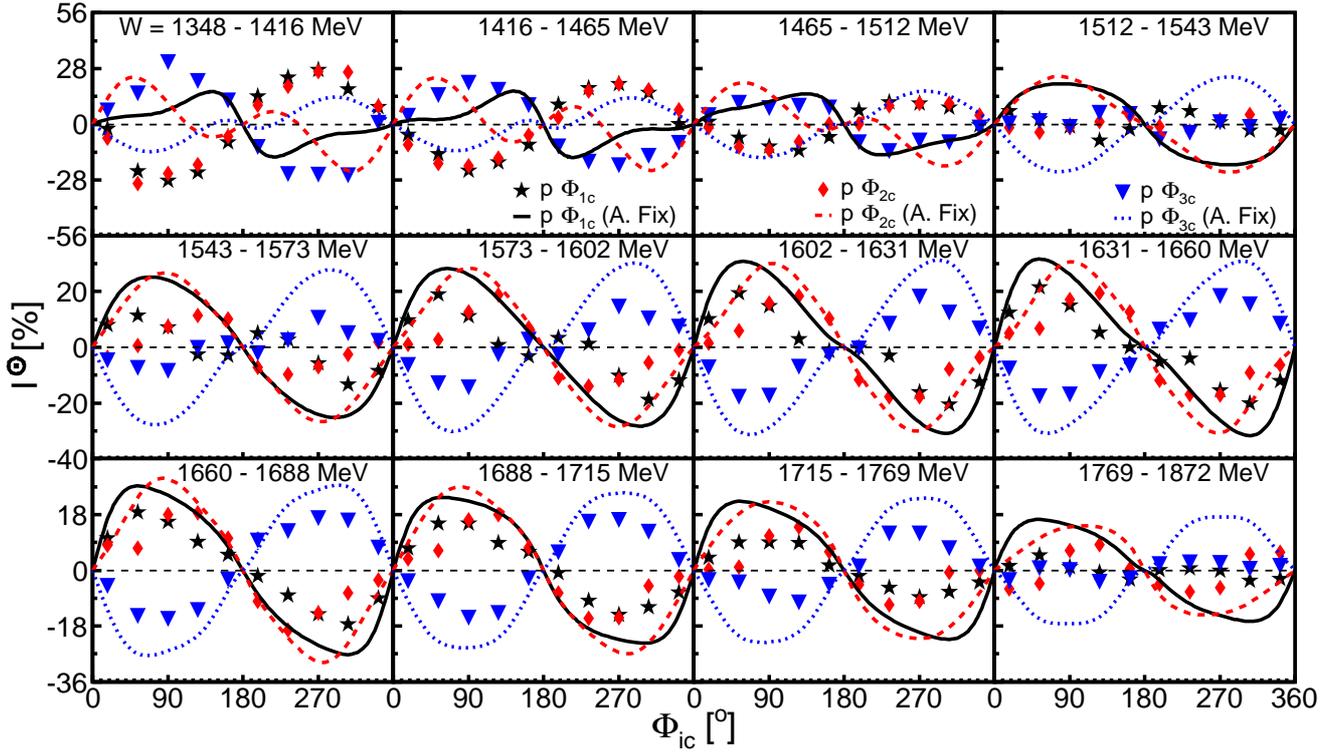}
}}
\caption{Results for $I^{\odot}_{1c}(\Phi_{1c})$ (black stars, black solid lines), 
$I^{\odot}_{2}(\Phi_{2})$ (red diamonds, dashed red lines)), and $I^{\odot}_{3}(\Phi_{3})$ 
(blue triangles, blue dotted lines) for the $\gamma p\rightarrow n\pi^0\pi^+$ reaction.
The symbols represent the data, the lines are the predictions from the Two-Pion-MAID model
\cite{Fix_05}. 
}
\label{fig:all_p}       
\end{figure*}

\begin{figure*}[!t]
\centerline  {\resizebox{0.99\textwidth}{!}{%
\includegraphics{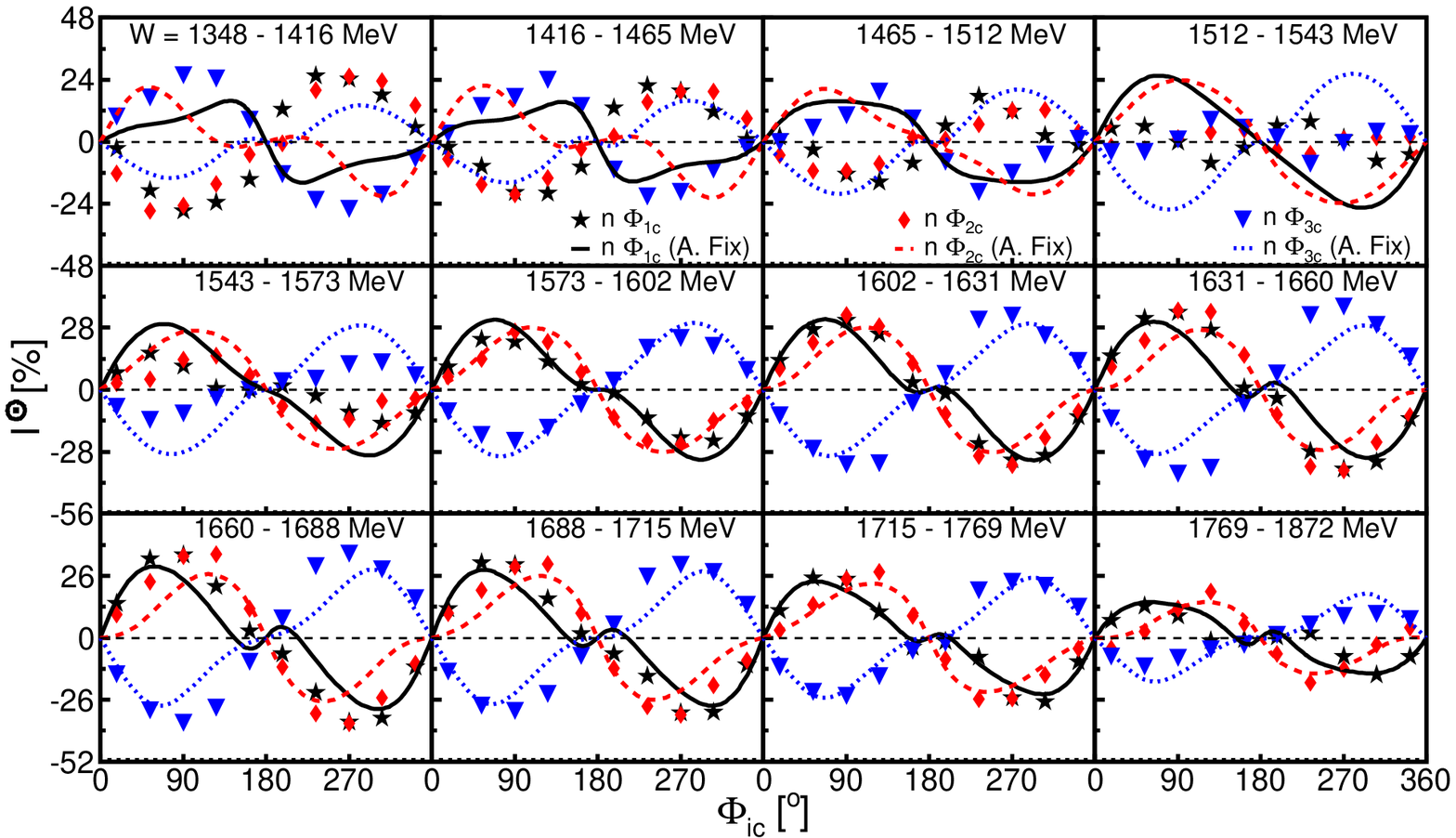}
}}
\caption{Same as Fig.~\ref{fig:all_p} for the $\gamma n\rightarrow p\pi^0\pi^-$ reaction. 
}
\label{fig:all_n}       
\end{figure*} 

In the following we summarize the most relevant results from the large body of data obtained
by the present experiment. The asymmetries $I^{\odot}_{1c}(\Phi_{1c})$ and $I^{\odot}_{1m}(\Phi_{1m})$ 
for the `charge' and `invariant-mass' ordering of the pions for the free proton measured with the 
hydrogen target and the quasi-free protons and neutrons from the deuterium target are shown for 
the full range of measured photon energies  in Figs.~\ref{fig:integ_asym_pi} and \ref{fig:integ_asym_im}. 
The three different `charge ordered' asymmetries $I^{\odot}_{1c}(\Phi_{1c})$, $I^{\odot}_{2c}(\Phi_{2c})$ ,
and $I^{\odot}_{3c}(\Phi_{3c})$ corresponding to ($p_1,p_2,p_3)=(\pi^{\pm},\pi^0,N'$), $(\pi^{0},N',\pi^{\pm}$),
and $(\pi^{\pm},N',\pi^{0}$) are compared in Fig.~\ref{fig:all_p} (for $\gamma p\rightarrow n\pi^0\pi^+$)
and Fig.~\ref{fig:all_n} (for $\gamma n\rightarrow p\pi^0\pi^+$). These data are for quasi-free 
production from nucleons bound in the deuteron. 

\begin{figure}[thb]
\resizebox{0.50\textwidth}{!}{%
  \includegraphics{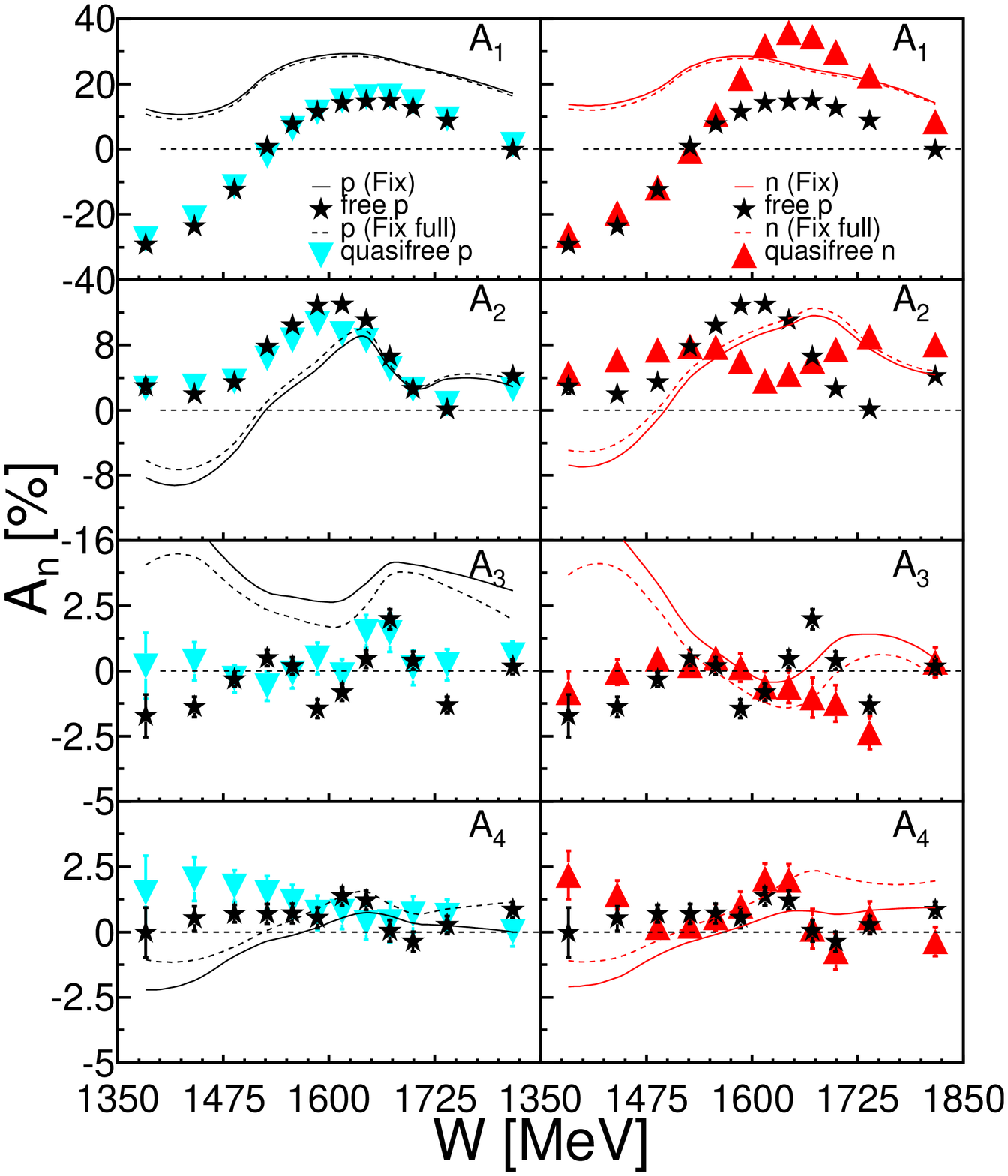}
}
\caption{Coefficients of the fits of the `charge-ordered' asymmetries $I^{\odot}_{1c}(\Phi_{1c})$
from Fig.~\ref{fig:integ_asym_pi} with Eq.~\ref{eq:coeff} as function of cm-energy $W$. 
Left hand side: free and quasi-free proton data, right hand side: comparison of proton and neutron 
asymmetries. 
Solid curves: model results from \cite{Fix_05} restricted to experimental acceptance. Dashed curves:
same model without acceptance restriction. 
}
\label{fig:nucl_par_pi}       
\end{figure}

All discussed asymmetries have been integrated over the
the full reaction phase-space with the exception of events where the charged pion was emitted to laboratory
polar angles smaller than 20$^{\circ}$ (i.e. into the angular range covered by TAPS).
The predictions from the Two-Pion-MAID model \cite{Fix_05} are compared in all figures to the data.

\begin{figure}[thb]
\resizebox{0.50\textwidth}{!}{%
  \includegraphics{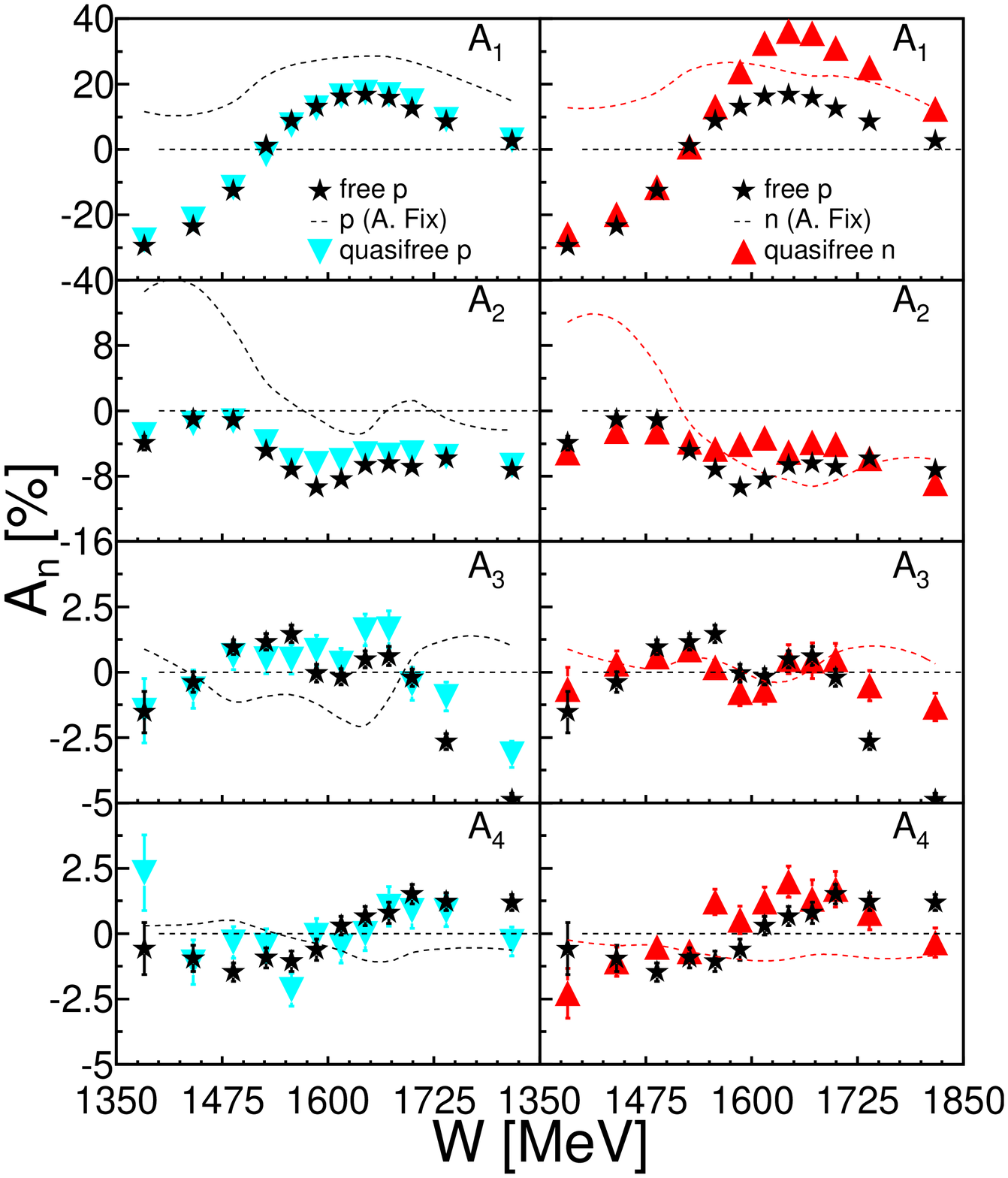}
}
\caption{Coefficients of the asymmetries $I^{\odot}_{2c}(\Phi_{2c})$. Notation as in
Fig.~\ref{fig:nucl_par_pi}.
}
\label{fig:para_2}       
\end{figure}

\begin{figure}[hb]
\resizebox{0.50\textwidth}{!}{%
  \includegraphics{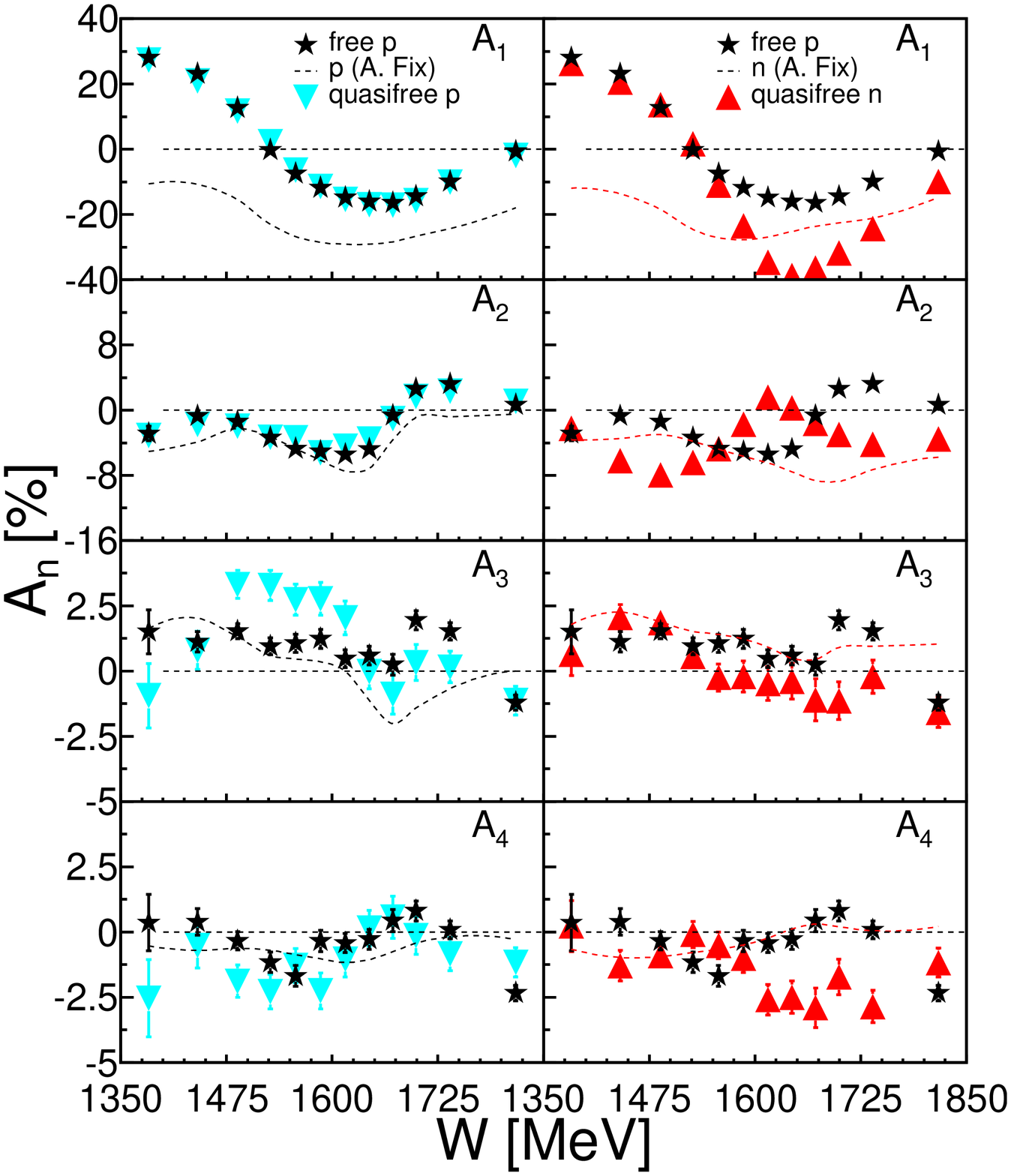}
}
\caption{Coefficients of the asymmetries $I^{\odot}_{3c}(\Phi_{3c})$. Notation as in
Fig.~\ref{fig:nucl_par_pi}. 
}
\label{fig:para_3}       
\end{figure}

\begin{figure}[thb]
\resizebox{0.50\textwidth}{!}{%
  \includegraphics{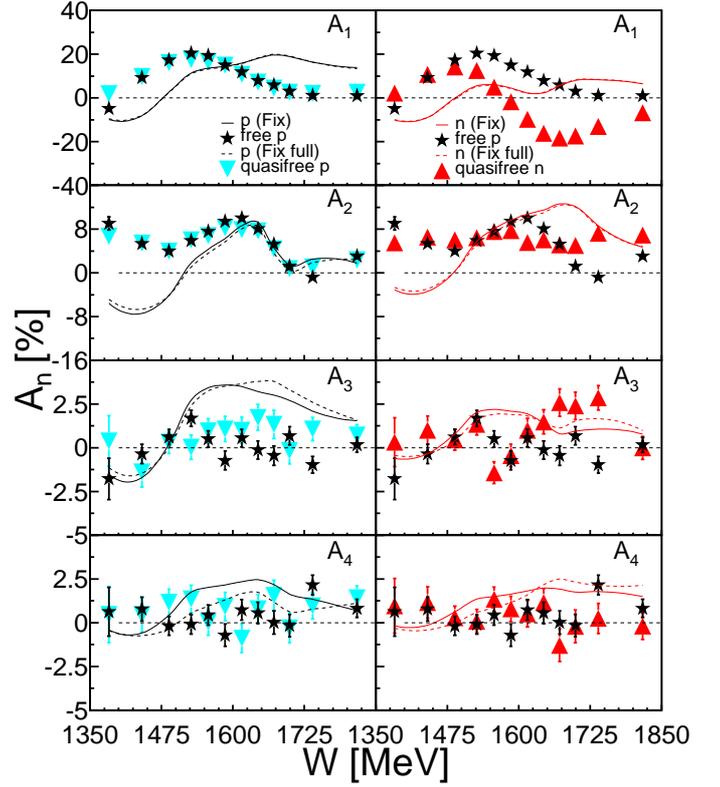}
}
\caption{Coefficients of the fits of the `mass-ordered' asymmetries $I^{\odot}_{1m}(\Phi_{1m})$
from Fig.~\ref{fig:integ_asym_im}. 
Notation as in Fig.~\ref{fig:nucl_par_pi}.
}
\label{fig:nucl_par_im}       
\end{figure}

The distributions have been fitted with the sine-series from Eq.~\ref{eq:coeff}. The results of the
fits for the coefficients are summarized and compared to the model predictions from \cite{Fix_05} in 
Figs.~\ref{fig:nucl_par_pi}-\ref{fig:nucl_par_im}. 

The free-proton data as well as the quasi-free proton and quasi-free neutron data were analyzed 
as a function of the final-state invariant mass $W$ (of the two-pion-participant-nu\-cleon system), 
which was reconstructed as discussed in Sec.~\ref{sec:ana}. All asymmetries for the free and 
quasi-free proton targets agree quite well, demonstrating that the kinematic reconstruction of the 
final state reliably eliminates the effects of nuclear Fermi motion (within experimental resolution). 
This is not trivial, Fermi motion modifies not only the effective $W$  but influences also the
orientation of the two planes and thus the angle $\Phi$.

Also this angle must be reconstructed for the `true' cm system. Analyses of the quasi-free data 
without careful reconstruction of the Fermi-motion related effects result in asymmetries with 
significantly reduced magnitudes. Further nuclear effects from final-state interactions 
(FSI), were not observed in the comparison of free and quasi-free proton data. This is the 
justification for the assumption that the quasi-free neutron data can be regarded as a close 
approximation of free-neutron data. The same observation has been previously made for the 
$\pi^0\pi^0$ final state \cite{Oberle_13}. FSI effects seem to be more important for absolute cross 
section data, however, even for such data they depend strongly on the reaction under study. 
Detailed model predictions for FSI are up to now only available for a few reaction channels.
Substantial effects have been found for the $\gamma n\rightarrow p\pi^-$ reaction measured with neutrons 
bound in the deuteron. Their energy and angular dependence has been studied with models beyond
the impulse approximation in references \cite{Tarasov_11,Chen_12}. Large effects have also been
found for the $\gamma N\rightarrow N\pi^0$ reactions off nucleons bound in the deuteron \cite{Dieterle_14}.
On the other hand, FSI effects for quasi-free photoproduction of $\eta$ 
\cite{Jaegle_08,Jaegle_11,Werthmueller_13} and $\eta'$-mesons \cite{Jaegle_11b} off nucleons from the 
deuteron are negligible, while for $^3$He nuclei also $\eta$-photoproduction shows large FSI 
\cite{Witthauer_13}.
 
The influence of the small acceptance restriction, which excluded charged pions in TAPS 
(i.e. at laboratory polar angles below  $20^{\circ}$) from the analysis was investigated 
by imposing the same restriction to the model results. 
Model results for $4\pi$ acceptance and for the restricted acceptance are shown
in Figs.~\ref{fig:integ_asym_pi},\ref{fig:integ_asym_im},\ref{fig:nucl_par_pi},\ref{fig:nucl_par_im}.
The results are so similar that this limitation is ignored in the further discussion. One should, 
however, take it into account when other model results are compared to the data.

\begin{figure*}[thb]
\resizebox{1.0\textwidth}{!}{%
  \includegraphics{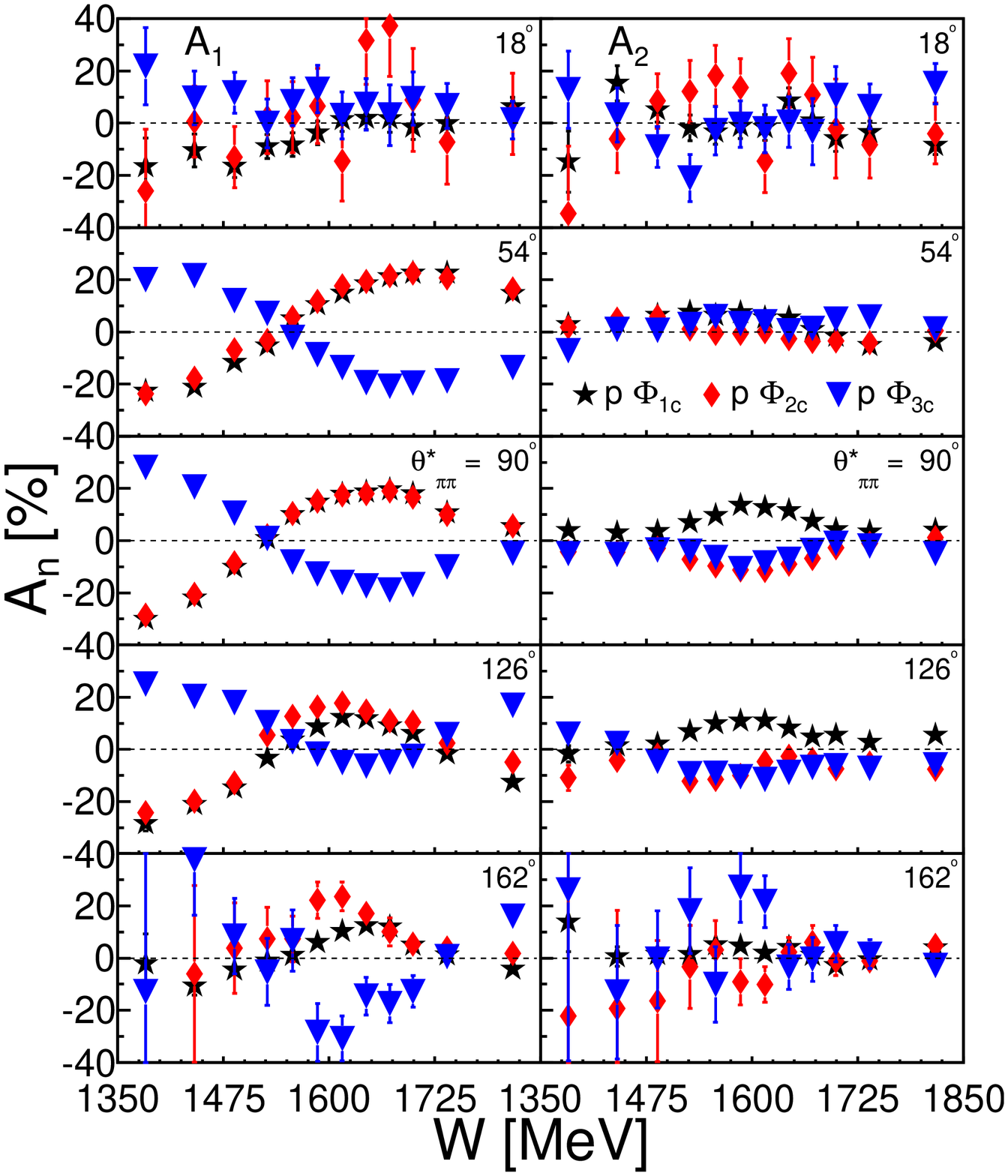}
  \includegraphics{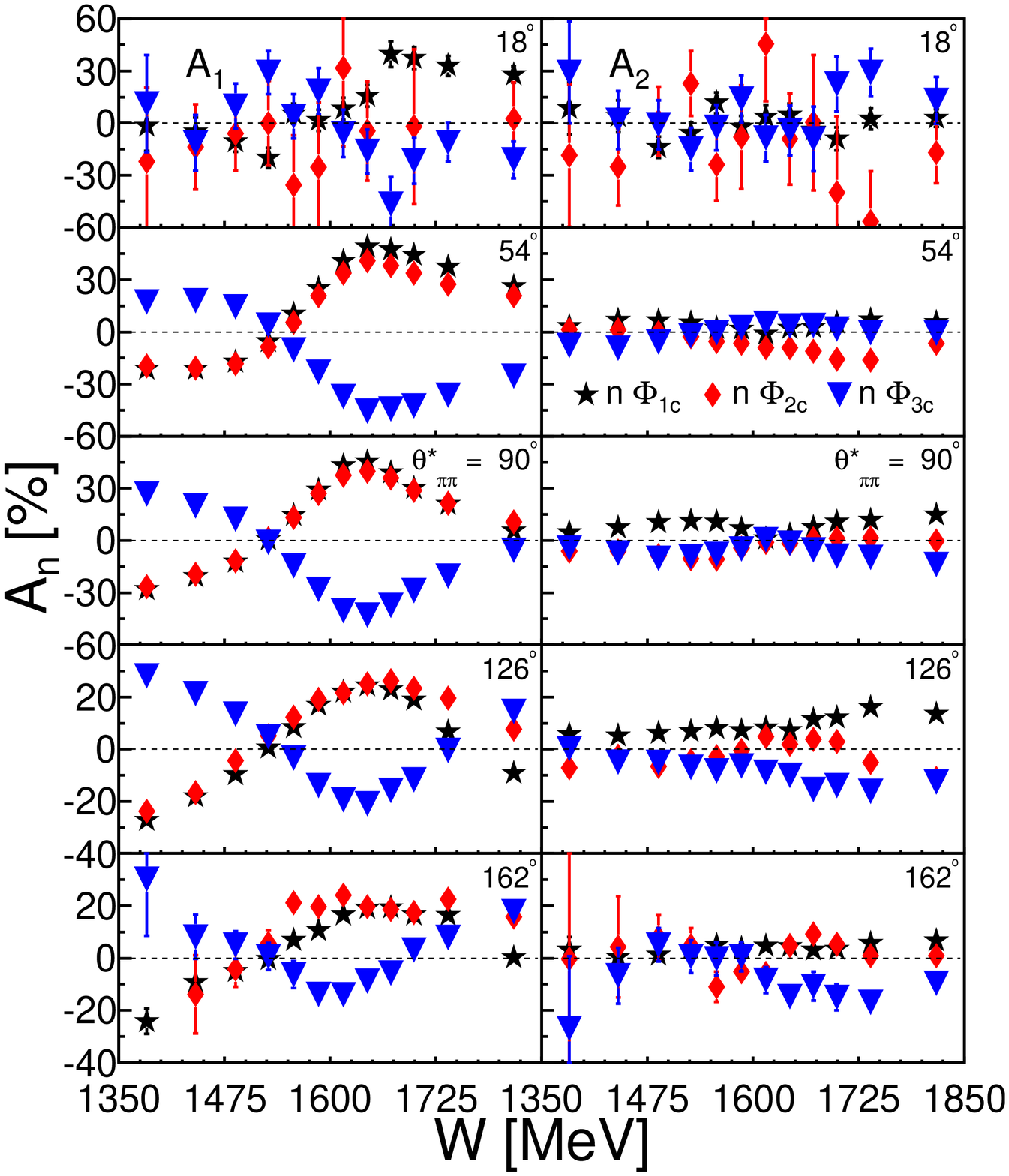}  
}
\caption{Coefficients $A_1$, $A_2$ for the asymmetries $I^{\odot}_{1c}(\Phi_{1c})$, 
$I^{\odot}_{2c}(\Phi_{2c})$, and $I^{\odot}_{3c}(\Phi_{3c})$ for different cm polar angles
of the $\pi\pi$ system as function of $W$ for the reactions $\gamma p\rightarrow n\pi^0\pi^+$ (left hand side)
and $\gamma n\rightarrow p\pi^0\pi^-$ (right hand side).
}
\label{fig:para_diff}       
\end{figure*} 

In the following we summarize the most prominent features of the data. All asymmetries are
dominated by the low-order terms $A_1$, $A_2$ of the sine-expansion from Eq.~\ref{eq:coeff}. 
Magnitudes up to 40\% are reached for $A_1$ (up to 10\% for $A_2$), while the higher orders 
are in the range of a few per cent and partly at the limit of statistical significance.

Although there is no generally valid relation between the three charge ordered 
asymmetries $I^{\odot}_{1c}(\Phi_{1c})$, $I^{\odot}_{2c}(\Phi_{2c})$, 
and $I^{\odot}_{3c}(\Phi_{3c})$ the present results seem to obey the relation 
\begin{equation}
A_1(I^{\odot}_{1c}) \approx A_1(I^{\odot}_{2c}) \approx -A_1(I^{\odot}_{3c})
\label{eq:a1rel}
\end{equation}
for the leading $A_1$ coefficient. This is true for the experimental results and also for the model  
predictions (although they are not in agreement with each other for the actual values of
the coefficients).
    
Due to its symmetry, the even coefficients of the sine-series should not depend on
the ordering of $p_1$ and $p_2$ i.e. they should be identical for $I^{\odot}_{1c}(\Phi_{1c})$ 
(Fig.~\ref{fig:nucl_par_pi}) and $I^{\odot}_{1m}(\Phi_{1m})$ (Fig.~\ref{fig:nucl_par_im})
and also for a random ordering of $p_1$, $p_2$. The odd coefficients depend on the ordering and 
have to vanish for random ordering. The latter condition was fulfilled within statistical
uncertainties. The small $A_4$ coefficient agrees for $I^{\odot}_{1c}(\Phi_{1c})$ and
$I^{\odot}_{1m}(\Phi_{1m})$ basically within statistical uncertainties. 
For the $A_2$ coefficient there are deviations between the two ordering schemes in particular
for protons at the lowest $W$ values. Comparison of the two data sets gives an indication for 
systematic uncertainties. Here one should note, that at these $W$ values $I^{\odot}_{1c}(\Phi_{1c})$
is strongly dominated by the $A_1$ term, which almost vanishes for $I^{\odot}_{1m}(\Phi_{1m})$ 
so that probably the fit results for the smaller coefficients at low $W$ are more reliable for
$I^{\odot}_{1m}(\Phi_{1m})$.

The results for the proton and neutron asymmetries are similar for $W$ below $\approx$1540~MeV,
i.e. in the second resonance region, but differ significantly for larger $W$, in particular 
for the invariant mass ordering of the asymmetries. In this respect they behave differently from 
the previously studied $\gamma N\rightarrow n\pi^0\pi^0$ reaction \cite{Oberle_13} for which 
the proton and neutron asymmetries agreed over the full energy range. The behavior observed here 
is more in line with expectations than the $N\pi^0\pi^0$ results. In the second resonance region 
states like the D$_{13}$(1520) are excited with comparable strength on protons and neutrons,
which may explain the similarities. However, at larger $W$, in the third resonance region, the
dominant resonance contributions for protons and neutrons come from different states, so that
different asymmetries are to be expected.

All results discussed so far were integrated over all kinematic variables apart from the final
state invariant mass. In Fig.~\ref{fig:para_diff} one example for differential results is shown.
Plotted are the leading $A_1$ and $A_2$ coefficients for the proton target as function of $W$
for different bins of the cm polar angle $\Theta^{\star}_{\pi\pi}$ of the pion-pion system 
(the cm polar angle of the recoil nucleon is $\Theta^{\star}_N = 180^{\circ}-\Theta^{\star}_{\pi\pi} $). 
The asymmetries must vanish for $\Theta^{\star}_N = 0^{\circ}, 180^{\circ}$ because in this cases the 
recoil nucleon and the incident photon are colinear so that no reaction plane is defined. The results 
for the asymmetries are of course statistically dominated by the values around 
$\Theta^{\star}_{\pi\pi} = 90^{\circ}$ (which is one of the reasons why the detection efficiencies 
cancel even in the ratio of the integrated asymmetries). Also for the differential results for proton 
and neutron the $A_1$ coefficients seem to be approximately related by Eq.~\ref{eq:a1rel}. A prominent
feature of all three asymmetries is the zero crossing of the $A_1$ coefficients at $W$=1525~MeV,
which is not reproduced by the model (see. Figs.~\ref{fig:nucl_par_pi},\ref{fig:para_2},\ref{fig:para_3}).

In the following we compare the measured asymmetries to the results from reaction models. The 
Valencia model \cite{Nacher_02,Roca_05} reproduced many features of this reaction
in the second resonance region (total cross sections, invariant-mass distributions, split
into $\sigma_{1/2}$ and $\sigma_{3/2}$ components of the cross section) reasonably well but failed 
for the beam-helicity asymmetry in the second resonance region \cite{Krambrich_09}.
Predictions for higher incident photon energies or for the neutron target are not available from 
this model. Predictions from the Bonn-Gatchina (BnGn) model, which described the $p\pi^0\pi^0$
data quite well \cite{Oberle_13}, are also not yet available for this isospin channel.
However, such analyses are now under way. The Bonn-Gatchina group has recently extended their
coupled channel analysis \cite{Anisovich_10} to the neutron target \cite{Anisovich_13}
and is currently including further reaction channels into the model.  
 
The Two-Pion MAID model \cite{Fix_05} was in reasonable agreement with the beam-helicity asymmetries 
for neutral pion pairs \cite{Oberle_13,Krambrich_09} in the second resonance region, although it reproduced 
total cross sections at low incident photon energies \cite{Zehr_12} not so good. However, it also failed 
for the asymmetries of the mix-charged pairs \cite{Krambrich_09} in that energy region. Currently this model 
is the only one that made predictions up to higher incident photon energies and for reactions off the proton 
and off the neutron, which are compared in the figures to the measured asymmetries. 

We discuss first the three `charge-ordered' asymmetries $I^{\odot}_{1c}, I^{\odot}_{2c}, I^{\odot}_{3c}$.
The result is somewhat surprising. The results from the model are at least in reasonable agreement
with the experimental findings for final-state invariant masses above 1550~MeV 
(cf Figs.~\ref{fig:all_p},\ref{fig:all_n}). However, as already discussed in \cite{Krambrich_09}
for $I^{\odot}_{1c}$, they disagree with the data and are out-of-phase for smaller 
$W$ in the second resonance region. One would expect that at these low energies, with only a small 
set of well-known nucleon resonances contributing, the model should perform more reliably than at
higher energies, but the contrary is the case.

The situation for the `invariant-mass ordered' asymmetries $I^{\odot}_{1m}$
(Figs.~\ref{fig:integ_asym_im},\ref{fig:nucl_par_im})
is even worse. Predictions for this observable fail in the second and in the third resonance region.
Only at intermediate $W$ (1550 - 1650 MeV) they are similar to the data, which indicates that also for
large $W$ the dynamics of the reaction mechanism is not completely understood. 
 
\section{Summary and conclusions}
Precise results have been measured for the first time for several types of beam-helicity asymmetries 
in the production of $\pi^0\pi^{+/-}$ pairs off free protons from a hydrogen target and off 
quasi-free protons and neutrons from a deuterium target with a circularly polarized photon beam.
Together with the results published in \cite{Oberle_13} for the $\gamma N\rightarrow N\pi^0\pi^0$
reaction and for $\gamma p\rightarrow p\pi^+\pi^-$ \cite{Strauch_05} such asymmetries are now
available for five different isospin channels of double pion production.

The asymmetries are sizable and the results for free and quasi-free protons are in excellent agreement
when the kinematics of the quasi-free reaction are completely reconstructed. This justifies the interpretation
of the quasi-free data for photoproduction off neutrons as a close approximation of free
neutron data.

At present only one reaction model, the Pion-MAID model \cite{Fix_05}, has made predictions for both 
reactions over the full energy range, but further model analyses are under way.
For the comparison between data and model results one can roughly distinguish three different ranges of
final state invariant mass $W$. These are the second resonance peak ($W\le$1540~MeV), the third resonance peak
($W\ge$1660~MeV), and the region in between. 

The analysis of the $\gamma N\rightarrow N\pi^0\pi^{\pm}$ reaction has a difficult history for the 
second resonance region \cite{Krusche_03}. Early measurements of total cross sections and invariant 
mass distributions did not agree with any model predictions. Agreement became better when significant 
contributions from the $\rho$-meson were introduced into the models, but the previous experimental 
results for the asymmetry $I^{\odot}_{1c}$ for free protons \cite{Krambrich_09} did again not agree 
with model predictions. The present results show that for all four considered asymmetries in this 
energy range the experimental data for proton and neutron targets are very similar but the model 
predictions are in all cases completely out of phase with them. Similar discrepancies have been 
reported in \cite{Krambrich_09} for the Valencia model \cite{Roca_05}. This is a strong indication 
that the reaction mechanisms for the second resonance region are simply not yet understood. 
The situation is different for the photoproduction of neutral pion pairs \cite{Krambrich_09,Oberle_13}. 
In that case good agreement of the asymmetry  $I^{\odot}_{1m}$ (other asymmetries are not available) 
with model predictions was found in the second resonance region for protons and neutrons. This suggests
that the problems for the mixed-charge channel are not due to the contributions of sequential resonance 
decays (which contribute to the production of $\pi^0\pi^0$ and $\pi^0\pi^{\pm}$ according to the 
respective Clebsch-Gordan coefficients) but to contributions that are not present for the $N\pi^0\pi^0$ 
final state. These are in particular processes involving the $\rho$-meson or non-resonant terms in 
charged pion production (pion-pole terms, $\Delta$-Kroll-Ruderman etc.).

The poor agreement between experimental data and results from reaction models for asymmetries 
involving production of pion pairs with at least one charged pion \cite{Strauch_05,Krambrich_09} 
raised the question whether these observables are well suited for the study of nucleon resonances 
or are possibly too sensitive to interference terms with small background contributions. But the 
picture that emerges from the present results is somewhat different. The asymmetries predicted 
for the intermediate energy range (1540~MeV$\le W\le$1660~MeV) are in much better agreement with 
the measurements. Here, it does not look like a severe discrepancy but more like the necessity 
for some fine tuning of the model. Also in the third resonance region the `charged-ordered' 
asymmetries are already in reasonable agreement with measurements, although the `mass-ordered' 
asymmetry is not yet reproduced. However, one should note that all asymmetries have been
predicted by a reaction model that was not fitted to data but used only input for nucleon resonance
parameters from the Particle Data Group. When more observables become available for the double-pion
photoproduction those parameters could of course be fitted to the data.  

The main conclusion is therefore that a specific problem for the production of `mixed-charge' 
pairs in the second resonance peak persists, while for higher final-state invariant masses the 
model predictions are already reasonable when one considers that up to now they could only be 
tested versus differential cross section data and that a `complete experiment' for double-pion 
production would require the measurement of at least 15 observables. Analyses of the data in 
the framework of further reaction models are necessary and under way and further observables 
(invariant mass distributions of pion-pion and pion-nucleon pairs, target asymmetry $T$, and double 
polarization asymmetries $E$ and $F$) have already been measured and are under analysis.  

\vspace*{0.5cm}
{\bf Acknowledgments}

We wish to acknowledge the outstanding support of the accelerator group 
and operators of MAMI. 
This work was supported by Schweizerischer Nationalfonds, Deutsche
Forschungsgemeinschaft (SFB 443, SFB/TR 16), DFG-RFBR (Grant No. 05-02-04014),
UK Science and Technology Facilities Council, STFC, European Community-Research 
Infrastructure Activity (FP6), the US DOE, US NSF and NSERC (Canada).

\end{document}